\documentclass[twocolumn,trackchanges,floatfix]{aastex62} 

\newcommand{\RNum}[1]{\uppercase\expandafter{\romannumeral #1\relax}}

\newcommand{\bef}{\begin{figure}}      
\newcommand{\eef}{\end{figure}}      
\newcommand{\bea}{\begin{eqnarray}}    
\newcommand{\eea}{\end{eqnarray}}      
\newcommand{\be}{\begin{equation}}      
\newcommand{\ee}{\end{equation}}  

\usepackage{lineno}
\graphicspath{{./}{figures/}}

\received{???}
\revised{???}
\accepted{???}
\shorttitle{Mass models of the Milky Way from the GAIA DR3 data-set}
\shortauthors{Sylos Labini et al.}
\usepackage{amsmath}

\begin{document}

\title{Mass models of the Milky Way and estimation of its mass from the GAIA DR3 data-set}
  
\author{Francesco  Sylos Labini}
\affil{Centro  Ricerche Enrico Fermi, Via Pansiperna 89a, 00184 Rome, Italy}
\affil{Istituto Nazionale Fisica Nucleare, Unit\`a Roma 1, Dipartimento di Fisica, Universit\'a di Roma ``Sapienza'', 00185 Rome, Italy}

\author{\v{Z}ofia Chrob\'akov\'a}
\affil{Faculty of Mathematics, Physics, and Informatics, Comenius University, Mlynsk\'a dolina, 842 48 Bratislava, Slovakia}

\author{Roberto Capuzzo-Dolcetta}
\affil{Dipartimento di Fisica, Universit\'a di Roma ``Sapienza'', 00185 Rome, Italy}
\affil{Centro  Ricerche Enrico Fermi, Via Pansiperna 89a, 00184 Rome, Italy}

\author{Mart\'in L\'opez-Corredoira}
\affil{Instituto de Astrof\'\i sica de Canarias, E-38205 La Laguna, Tenerife, Spain}
\affil{Departamento de Astrof\'\i sica, Universidad de La Laguna, E-38206 La Laguna, Tenerife, Spain}

\correspondingauthor{FSL}
\email{sylos@cref.it}

\begin{abstract}
We use data from the Gaia DR3 dataset to estimate the mass of the Milky Way (MW) by analyzing the rotation curve in the range of distances 5 kpc to 28 kpc.  
We consider three mass models:  the first model adds a spherical dark matter (DM) halo, following the Navarro-Frenk-White (NFW) profile, to the known stellar components.
The second model assumes that DM is confined to the Galactic disk, following the idea that the observed density of gas in the Galaxy 
is related to the presence of more massive DM disk (DMD), similar to the observed correlation between DM and gas in other galaxies.
 The third model only uses the known stellar mass components and is based on the Modified Newton Dynamics (MOND) theory.
{ Our results indicate that the DMD model is comparable in accuracy to the NFW and MOND models and fits the data better at large radii where the rotation curve declines but has the largest errors.}
 For the NFW model we obtain a virial mass $M_{vir}=  (6.5 \pm 0.3)  \times 10^{11} \; M_\odot$ with concentration parameter $c=14.5$,  that 
is lower than what is typically reported.  In the DMD case we find that the MW  mass is  $M_d =  (1.6 \pm 0.5)  \times 10^{11} \; M_\odot$ with a disk's characteristic 
radius of $R_d=17$ kpc. 
\end{abstract}

\section{Introduction} 

Determining the Milky Way (MW)'s mass profile requires measuring its mid-plane circular velocity $v_c(R)$.
The  rotation curve has been measured  using different  methods 
and  kinematical data on a
variety of tracer objects (see, e.g., 
\cite{Bhattacharjee_etal_2014,Sofue_2020} and references therein).
However, in most cases, the full three-dimensional velocity information of the tracers is not available, so the circular velocity had to be estimated using only the measured line-of-sight velocity and position.
Uncertainties in distance estimates, limited numbers of tracers, and their uneven distribution can introduce significant errors in the analysis of the circular velocity curve.

To accurately determine the  MW's rotation curve $v_c(R)$, we need precise measurements of the Galactocentric radius
$R$, tangential velocity, and radial velocity for each star, including the position and velocity
uncertainties
 in all three spatial dimensions. 
The Gaia mission \citep{Gaia_2016}, by measuring the astrometry, photometry, and spectroscopy of a large number of stars, is providing the position and velocity information in all six dimensions for a large sample of stars in the MW.
These data are thus  ideal  to measure 
the Galaxy rotation curve $v_c(R)$. 

Recently, three independent research groups
\citep{Eilers_etal_2019,Mroz_etal_2019,Wang_etal_2023}
 have used the Gaia data-sets to determine the MW rotation curve, with measurements based on 
 different samples of stars. 
 The first two measurements are based on red giant stars and Cepheids, respectively, while the third uses a statistical deconvolution method applied to the entire data-set.
 These measurements show a similar slow declining trend in different but overlapping distance ranges between 5 kpc and 28 kpc.

It is clear that the mass estimated when the rotation curve declines 
is lower than that measured when $v_c(R) \approx$ const.
 Indeed,  
\cite{Eilers_etal_2019}, by considering the 
standard Navarro, Frenk and White (NFW) halo model \citep{Navarro_etal_1997}, found 
a virial mass 
of $M_{vir} = (7.25 \pm 0.25) \times 10^{11} M_\odot$ 
(with $R_{vir}=(189.3\pm2.2)$ kpc) 
which is significantly lower than what
several previous studies suggest (see, e.g., \cite{Watkins_etal_2019}),
although values reported in the literature span  approximately in the range 
$(0.5-3) \times 10^{12} M_\odot$ (see, e.g., \cite{Bland-Hawthorn_etal_2016} and references therein). 

In this paper we combine the determination of the rotation curve derived by
\cite{Eilers_etal_2019}, in the range of Galactocentric radii $5-25$ kpc, with that 
by \cite{Wang_etal_2023}, in the range $8-28$ kpc, 
to make an estimation of the mass of the MW under
three different theoretical mass models. 
The first theoretical mass model we are using is the canonical NFW halo model \citep{Navarro_etal_1997}, which assumes that the visible matter of the 
MW is confined to a rotationally supported disk embedded in a much heavier halo of dark matter (DM).
As the study by \cite{Wang_etal_2023} found that the declining trend in the rotation curve continues at greater distances, we expect that this mass model would predict a lower value of the MW's mass than the one estimated by \cite{Eilers_etal_2019}.

The second theoretical mass model assumes that 
DM  is confined to a relatively thin disk, similar to the visible disk.
The motivation for this model comes from the "Bosma effect" 
  \citep{Bosma_1978,Bosma_1981} which suggests a correlation between DM and gas in disk galaxies.
 Indeed, there is  substantial observational evidence that  rotation curves of disk galaxies are, at large enough radii, a re-scaled version of 
 those  derived from   the distribution of gas \citep{Sancisi_1999,Hoekstra_etal_2001}. 
It is important to note that the correlation between gas and DM observed in disk galaxies does not necessarily imply causality.
It is worth exploring the possibility that by assuming that the distribution of gas is a tracer of the distribution of DM, 
we can find a different mass model than the standard NFW one that fits the observed rotation curve of the MW  with similar accuracy, 
as it has been observed in a number of external galaxies by 
\cite{Hessman+Ziebart_2011,Swaters_etal_2012}.
In the DMD model, where DM is assumed to be confined to a relatively thin disk similar to the visible disk, 
the total mass of the Galaxy is expected to be lower than in the case of a spherical halo, 
as the mass is concentrated in a thinner region.

The third theoretical mass model is based on the framework of Modified Newtonian Dynamics (MOND) \citep{Milgrom_1983,scarpa06,McGaugh_etal2016}. This model does not require the introduction of an additional DM component and only considers the mass of the visible stellar components.
In this model, the gravitational force is assumed to decline more slowly than in Newtonian dynamics, which allows for the rotation curve to remain steady without the need for additional mass beyond the visible stellar components.

The paper is organized as follows. In Sect.\ref{bosmaeffect}, we discuss the main characteristics of the "Bosma effect" and its observational evidence. In Sect.\ref{rotationcurve}, we briefly review the measurements of the rotation curve based on the Gaia data-sets that we use in this work. In Sect.\ref{massmodels}, we discuss the different mass models and present their fits to the rotation curve. Finally, in Sect.\ref{conclusions}, we discuss the results obtained, draw our main conclusions, and mention the possible dynamical implications.


\section{The Bosma effect} 
\label{bosmaeffect} 

\cite{Bosma_1978,Bosma_1981} 
first noticed a correlation between the centripetal contribution of the atomic hydrogen H\RNum{1} gas, which is dynamically insignificant, in the disks of spiral galaxies and the dominant contribution of DM.
This correlation is known as the "Bosma effect" and has been observed by multiple studies \citep{Sancisi_1999,Hoekstra_etal_2001}. The effect is thought to reveal a relationship between the visible baryonic matter and the invisible DM.
For example, \cite{Hoekstra_etal_2001} examined a sample of disk galaxies with high-quality rotation curves and found that, with a few exceptions, the rotation curves generated by scaling up the centripetal contribution of the H\RNum{1} gas by a constant factor of about 10 and not including a spherical DM halo were similar in accuracy to those generated by the NFW halo model.

It should be noted that \cite{Hoekstra_etal_2001} defined the constant of proportionality between the centripetal effects of the H\RNum{1} gas and the DM as the assumed constant ratio of the DM to H\RNum{1} surface densities, averaged over the disk. This ratio was corrected only for the presence of helium, which is a small fraction of the total gas mass.
Bosma's original concept was to use the total gas surface density as a proxy for DM, including not only H\RNum{1} but also other components of the interstellar medium (ISM).
\cite{Hessman+Ziebart_2011} distinguished between "simple" Bosma effect, which is the case where only the surface density of H\RNum{1} (corrected for the contribution of He and heavy elements) is used as a proxy for the distribution of DM, and "classic" Bosma effect, which includes the total gaseous surface density, not only of H\RNum{1} but of other components of the ISM. 
This can be done explicitly by using 
the surface density of the ISM defined as 
$\Sigma_{ISM} = \Sigma_{HI} + \Sigma_{H_2}$ 
 (again corrected for He and heavy elements), or implicitly 
by using the stellar disk as an additional proxy. 

The reason for this inclusion is that, from a physical point of view, it is expected that the correlation between DM and the ISM would be with the total ISM, not just its neutral hydrogen (H\RNum{1}) component. This is because the total ISM, including both neutral and ionized gas, is thought to be more closely related to the distribution of DM than just the neutral H\RNum{1} component alone.

 \cite{Hessman+Ziebart_2011} used the stellar disk as an additional proxy of the DM component, because the H\RNum{1} surface density does not reflect the total gas surface density in the inner galactic region. They confirmed the correlation between DM and H\RNum{1} distribution using several galaxies from The H\RNum{1} Nearby Galaxy Survey dataset \citep{deBlok_etal_2008}  and rebutted several arguments against the effect by  \cite{Hoekstra_etal_2001}. They found fits of similar or even better quality than those obtained by the standard NFW halo model.
 
\cite{Swaters_etal_2012} further studied a sample of 43 disk galaxies by fitting the rotation curves with mass models based on scaling up the stellar and H\RNum{1} disks. They found that such scaling models fit the observed rotation curves well in the vast majority of cases, even though the models have only two or three free parameters (depending on whether the galaxy has a bulge or not). They also found that these models reproduce some of the detailed small-scale features of rotation curves such as the "bumps" and "wiggles".

In summary, the "Bosma effect" implies a close connection between the ISM and the DM and points towards a baryonic nature and a more or less flat distribution of the dark component. This could be in the form of very dense cold gas distributed in molecular clouds in galactic disks, as suggested by studies such as 
\cite{Pfenniger_etal_1994} and  \cite{Revaz_etal_2009}.


\section{Rotation curve of the Milky Way} 
\label{rotationcurve} 

\subsection{The data} 

Three independent determinations of the MW  rotation curve have been obtained from different samples based on the Gaia data, which reasonably agree with each other. These samples cover the range of distances between 5 kpc and 28 kpc, but each of them only partially. In the overlapping range of radii, they show reasonable agreement. The common feature is that the rotation curve, $v_c(R)$, presents a declining behavior with the cylindrical radius, $R$.

The first analysis was provided by \cite{Mroz_etal_2019}, who built and analyzed a sample of 773 Classical Cepheids with precise distances based on mid-infrared period-luminosity relations, coupled with proper motions and radial velocities from Gaia, in the range of radii between 5 kpc and 20 kpc. However, the number of Cepheids significantly decreases for $R>15$ kpc, which limits the well-sampled range of distances to 15 kpc.
They found that 
\be
\label{VC} 
v_c(R)= v(R_\odot)  + \beta (R - R_\odot) 
\ee
{  
where $R_\odot  = 8.122 \pm 0.031$ kpc  \citep{Gravity_collaboration_2018} 
is the distance of the Sun from the 
Galactic center, 
}
$v(R_\odot)$  the rotation speed of the Sun 
and
the  slope  was determined to be 
$\beta=-(1.4 \pm 0.1)  \;  \mbox{km s}^{-1} \mbox{kpc}^{-1}$. 

The second measurement was made by 
\cite{Eilers_etal_2019}  by 
using the spectral data from APOGEE
DR14, the  photometric information from Gaia Data Release 2 (DR2), 2MASS,
and the Wide-field Infrared Survey Explorer. They  
built a sample of 23,000 red giant stars with the 
whole  6D spatial 
and velocity information.
The sample is thus  characterized by  
precise and accurate distances to
luminous tracers that can be observed over a wide range of
Galactic distances. 
They  derived the 
rotation curve 
for   Galactocentric distances between 
5 kpc and 25 kpc and 
found a slope 
$\beta=-(1.7 \pm 0.1)  \;  \mbox{km s}^{-1} \mbox{kpc}^{-1}$
that  is in reasonably good agreement with 
that of  \cite{Mroz_etal_2019} considering the 
former covers the range $\approx 5-15$  kpc while 
the latter $5-25$ kpc. 

The third determination was made by  
\cite{Wang_etal_2023} 
who have adopted the  statistical  inversion method introduced by \cite{Lucy_1977}
to reduce the  errors in the distance determination
in  the Gaia DR3 data-set beyond 20 kpc. That method 
was firstly applied by 
\citet{Lopez-Corredoira_Sylos-Labini_2019}   
 to the Gaia DR2 sources 
to measure their three dimensional velocity components in 
the range of Galactocentric distances  between 
8 kpc and 20 kpc with their corresponding errors and root mean square values.
\cite{Wang_etal_2023} have extended the analysis  to $\approx 28$ kpc  by considering 
the Gaia DR3 sources \citep{Gaia_DR3} and they have obtained results 
for the three velocity components in agreement with those 
measured by \cite{Lopez-Corredoira_Sylos-Labini_2019}.
In addition, the rotation curve $v_c(R)$  obtained by \cite{Wang_etal_2023} 
is in reasonable agreement with the measurements by 
\cite{Mroz_etal_2019} and \cite{Eilers_etal_2019}. 
In particular, the slope
of Eq.\ref{VC} was $\beta=-(2.3 \pm 0.2)  \; \mbox{km s}^{-1} \mbox{kpc}^{-1}$,
i.e. a faster declining slope  than the other two determinations mentioned above. 
However, in the range of radii
where they overlap, the three measurements are 
in reasonably good agreement with each other.

We use the  determinations by \cite{Wang_etal_2023}
with that of \cite{Eilers_etal_2019} to construct the rotation curve in  
the range $5-28$ kpc. We report in Tab.\ref{tabrot} the values 
of the rotation curve in bins of size $\Delta R=0.5$ kpc
and we label this rotation curve  DR3+ to
distinguish from the determinations of  \cite{Eilers_etal_2019}  (E19)   
and of  \cite{Wang_etal_2023} (DR3).
In what follows we will provide fits with different mass models 
for all these three measurements of the rotation curve.

\begin{table} 
\begin{center}
\begin{tabular}{ c c c }
\hline 
\hline 
 $R$ (kpc)  & $v_c$ (km s$^{-1}$)  &  $\sigma_{v_c}$ (km s$^{-1}$)   \\  
\hline 
   5.25    &    226.8 &      1.9  \\   
   5.75    &    230.8 &      1.4  \\ 
   6.25    &    231.2 &      1.4  \\ 
   6.75    &    229.9 &      1.4  \\  
   7.25    &    229.6 &      1.2 \\ 
   7.75    &    229.9 &      0.9  \\  
   8.25    &    228.9 &      0.7  \\  
   8.75    &    224.2 &      2.4  \\  
   9.25    &    223.8 &      2.0  \\  
   9.75    &    223.7 &      2.2  \\  
   10.25   &    224.5 &      1.8  \\  
   10.75   &    223.6 &      1.8 \\  
   11.25   &    222.2 &      2.0  \\  
   11.75   &    222.4 &      1.8  \\  
   12.25   &    222.0 &      1.9  \\  
   12.75   &    222.4 &      2.2  \\  
   13.25   &    223.1 &      1.8  \\  
   13.75   &    220.5 &      1.9  \\  
   14.25   &    219.5 &      2.3  \\  
   14.75   &    219.3 &      1.9  \\  
   15.25   &    218.0 &      2.1  \\  
   15.75    &   217.6 &      2.4  \\  
   16.25    &   216.4 &      2.4  \\  
   16.75    &   214.0 &      2.4  \\  
   17.25    &   215.1 &      2.8  \\   
   17.75    &   213.3 &      2.7  \\  
   18.25    &   208.9 &      2.7  \\  
   18.75    &   209.1 &      3.4  \\  
   19.25    &   206.9 &      3.2  \\  
   19.75    &   203.4 &      3.3  \\  
   20.25    &   201.2 &      3.5  \\  
   20.75    &   200.1 &      3.5  \\  
   21.25    &   197.3 &      5.3  \\  
   21.75    &   205.2 &      9.2  \\  
   22.25    &   184.0 &      13.5 \\  
   22.75    &   194.4 &      13.4 \\  
   23.25    &   195.2 &      4.5  \\  
   23.75    &   185.1 &      9.7  \\  
   24.25    &   196.3 &      1.7  \\  
   24.75    &   193.4 &      3.7  \\  
   25.25    &   190.3 &      10.4 \\  
   25.75    &   179.3 &      9.4  \\  
   26.25    &   174.3 &      9.2  \\  
   26.75    &   178.4 &      8.1  \\  
   27.25    &   176.9 &      8.2  \\
   \hline
\end{tabular}
\end{center}
\caption{
Measurements of the Circular Velocity of the Milky Way 
for the DR3+ rotation curve (see text).
Columns show the Galactocentric radius, the circular velocity, and its
 error bar.}
\label{tabrot} 
\end{table}

\subsection{Discussion} 

{  

The Galaxy rotation curve is obtained by using the time-independent Jeans equation in an axisymmetric gravitational potential and assuming a smooth and monotonic density distribution (see Eqs.(7)-(10) of \cite{Wang_etal_2023}). This basic assumption is used to link the moments of the velocity distribution and the density of a collective of stars to the gravitational potential and to derive the Jeans equation and to derive the Jeans equation 
\citep{Binney_Tremaine_2008}. However, simplifications are often made, such as neglecting terms with $v_Z$ in the Jeans equation 
\citep{Eilers_etal_2019,Wang_etal_2023} and non-monotonic variations in the surface density profile, like localized structures like spiral arms, as these may cause deviation from the smooth models that are generally assumed  \citep{McGaugh_2016}. These structures show that the assumption of a time-independent gravitational potential is only a rough approximation to the actual dynamics.

The observed velocity field in the galaxy has been found to have asymmetrical motions with significant gradients in all velocity components
 \citep{Katz_etal_2018,Antoja_etal_2018,Lopez-Corredoira_Sylos-Labini_2019,Khoperskov_etal_2021,Wang_etal_2023}, which implies that the hypotheses used to obtain the Galaxy rotation curve, such as the time-independent Jeans equation and a smooth and monotonic density distribution, must be considered as approximations to the actual dynamics. Constructing a theoretical, self-consistent description of the galaxy that can relate these streaming motions in all velocity components to spatial structures is a challenging task. The gravitational influence of various galactic components, such as the long bar or bulge, spiral arms, or a tidal interaction with Sagittarius dwarf galaxy, may explain some features of the observed velocity maps, particularly in the inner parts of the disk. However, in the outermost regions, the main observed features can only be explained by out-of-equilibrium models, which are either due to external perturbers or to the fact that the disk has not had enough time to reach equilibrium since its formation \citep{Lopez-Corredoira_etal_2020}.

\cite{Chrobakova_etal_2020} found that, as long as the amplitude of the radial velocity component is small compared to that of the azimuthal one, the Jeans equation provides a reasonable approximation to the system dynamics. Kinematic maps derived from the Gaia DR3 data-set in the Galactic plane up to 30 kpc show that perturbations in the radial velocity are small enough so that using the time-independent Jeans equation to compare observations with theoretical models is justified \citep{Wang_etal_2023}. 

A standard practice in mass modeling of external galaxies is to numerically solve the Poisson equation for the observed surface brightness distribution, which allows for non-monotonic variations in the surface density profile and/or the presence of bumps and wiggles to be taken into account \citep{Sellwood_1999,Palunas+Williams_2000,deBlok_etal_2008}.
Specific pattern of bumps and wiggles in its density profile  should leave a distinctive imprint on the predicted rotation curve \citep{Sancisi_2004}.
 Both \cite{Eilers_etal_2019} and \cite{Wang_etal_2023} have assumed that the volume mass density of matter has an exponential decay.
 Here,  instead of assuming the exponential function, the surface stellar density profile was computed from the star distribution in the Gaia EDR3 data-set \citep{Chrobakova_etal_2022} and the logarithmic gradient of this profile was numerically computed. 
 This same method was used by \cite{McGaugh_2019} who determined at small radii, i.e. $R<8$ kpc, the influence
 of spiral arms on the rotation curve. 
 The observed stellar profile for $R>8$ kpc, reported in Fig.\ref{SB_density}, 
 does not have particular features and is well approximated by an exponential decay, which is why the analytical approximation works quite well and the differences with the observed one are smaller than the reported error bars.%
\begin{figure} 
\includegraphics[width=3.5in]{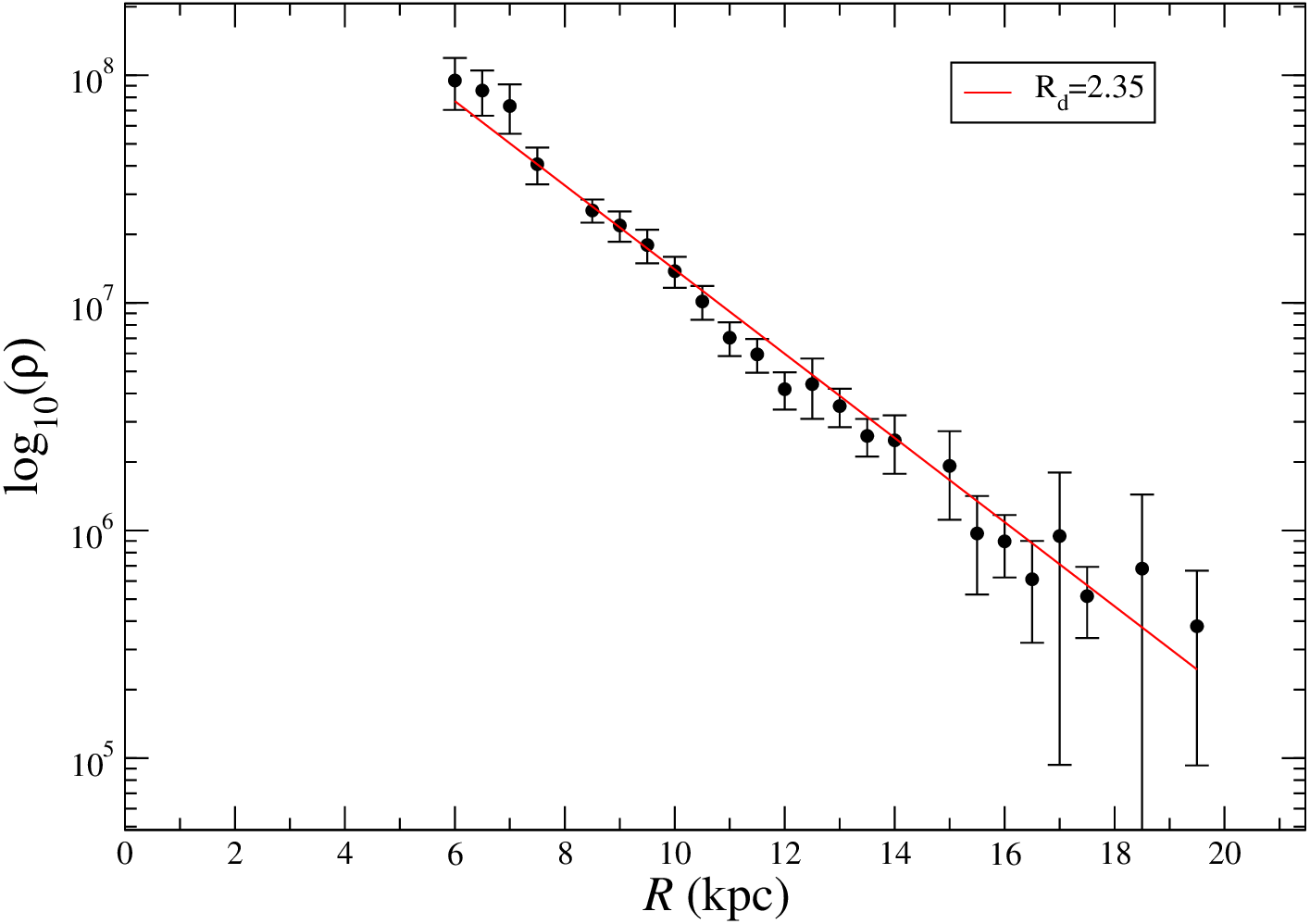}
\caption{
Dependence of the stellar density from the Gaia EDR3 data-set 
on the Galactocentric distance in the Galactic equatorial plane for 
the azimuth
$\phi \in [330^\circ; 30^\circ]$ 
(see  \cite{Chrobakova_etal_2022} for details).
}
 \label{SB_density} 
\end{figure}

Finally, in order to estimate systematic uncertainties on the circular velocity curve arising from the data sample, following \cite{Wang_etal_2023}, the galactic region was split into two disjoint smaller portions, one with galactic latitude $b > 0^\circ$ and the other with $b < 0^\circ$ or one with $Z>0$ and one with $Z<0$. The rotation curve was then computed in the two disjoint regions and the systematic uncertainties on the circular velocity were estimated by the difference between the resulting fit parameters from the two disjoint data sets. 
We find that the rotation curves obtained are within the reported error bars, leading to the conclusion that systematic fluctuations, such as those due to local structures, should not give a large contribution. With these hypotheses in mind, the rotation curve data can be used to fit different mass models.
}

%


\section{Mass Models} 
\label{massmodels} 

Given the profile of the rotation curve $v_c(R)$ 
presented in the last section and reported in  
Tab.\ref{tabrot}, 
we can now 
determine the parameters of the 
theoretical models.
Following  \cite{Eilers_etal_2019} we  assume 
that 
the stellar components consist of 
 the bulge, the thin disk
 and  the thick disk. We take the same characteristic values
 of these components as in \cite{Eilers_etal_2019}: the 
 final results do not qualitatively depend on this choice, but they
 do quantitatively. 
Because  Poisson's equation is linear 
both the gravitational potential and the 
square of the rotation $v_c(R)$ are a linear sum of the 
various components, i.e.   
\be 
\Phi= \sum_i \Phi_i \Rightarrow v_c^2= \sum_i v_{c,i}^2 \;.
\ee
The rotation velocity  $v_{c,i}$ is defined as  the velocity that the $i^{th}$ mass component would induce on a test particle in the plane
of the galaxy if it were placed in isolation without any external influences.
These velocities in the plane are calculated from the observed stellar mass density 
distributions plus the one due to 
the assumed DM distribution.

The first mass model considered in addition to the stellar components, takes into account the velocity contribution of the DM halo with a NFW profile \citep{Navarro_etal_1997}. This model has only two free parameters describing the NFW profile {  which are bound by a phenomenological relation
\citep{Maccio_etal_2008}}. The second mass model assumes the DM component to be distributed on the Galactic disk in agreement with the "Bosma effect". Two different scenarios are considered in this model. In the first case, DM is distributed on a thin disk with an exponentially decaying surface density profile, where the characteristic length of the profile $R_d$ and the total disk mass $M_d$ are considered as free parameters. In the second case, the functional behavior of the DM surface density is assumed to be the same as that of the observed Galactic gas distribution, where the length scale $R_d$ is that characterizing the gas surface density and the unique free parameter of the model is the disk total mass $M_d$.
Finally, the third mass model assumes only the stellar mass components, but in the MOND framework. In this case, the stellar disk characteristic length scale $R_d$ and mass $M_d$ are considered as free parameters.


\subsection{Disk and bulge components}
\label{stellar}

For the gravitational potentials of
the thin and thick disk 
\cite{Eilers_etal_2019}
used a  
Miyamoto-Nagai profiles
\cite{Miyamoto+Nagai_1975}, while  for the bulge they assume a
spherical Plummer potential \citep{Plummer_1911}. In addition, they adapted
the parameter values of the enclosed mass, the scale length, and
the scale height from \cite{Pouliasis_etal_2017} (model I).
We follow here this same approximation for the stellar components so that we find 
the thin disk has a characteristic length scale 
$R_{thin} \approx 4.5$ kpc and $M_{thin} \approx 3 \times 10^{10} M_\odot$,  
the thick disk has 
$R_{thick} \approx 2.3 $ kpc and $M_{thick} \approx 2.7 \times 10^{10} M_\odot$.
Models of the bulge 
gives a total bulge mass $M_{bulge}  \approx 2 \times 10^{10} M_\odot$  
with $R_{bulge} \approx 0.25 $ kpc 
\citep{Juric_etal_2008,Bland-Hawthorn_etal_2016}.
The characteristic length of the bulge, $R_{bulge} \approx 0.25 $ kpc, is small enough that the bulge 
contribution is relevant only at very small scales where the rotation curve is not well determined.
In addition, note that the  value of the mass of the bulge $M_{bulge}  \approx 2 \times 10^{10} M_\odot$   
is about twice larger than that used by \cite{Eilers_etal_2019}.

The total mass
of the stellar components, with these approximations, is 
$M_{stellar}  \approx 8 \times 10^{10} M_\odot$.


\subsection{The NFW halo  model}

The NFW mass model  can be written as 
\be
v_c^2 = v_{thin}^2 +  v_{thick}^2+ v_{bulge}^2+ v_{NFW}^2
\ee
where $v_{NFW}$ 
corresponds to the equilibrium velocity in a NFW profile described by
\citep{Navarro_etal_1997}
\be
\rho(r)= \frac{\rho_0}{\frac{r}{R_s} \left( 1 + \frac{r}{R_s} \right)^2} \;,
\ee
where $R_s$ and $\rho_0$ are two free parameters. Results can be expressed in terms of 
the virial radius, where the mean density of the halo reaches a value
200 times that of the mean cosmic mass-density, 
$R_{vir} = c R_s $
and  halo virial mass is $M_{vir} = M(R_{vir})$,
where $c$ is the  concentration parameter   \citep{Navarro_etal_1997}.

\begin{figure} 
\includegraphics[width=3.5in]{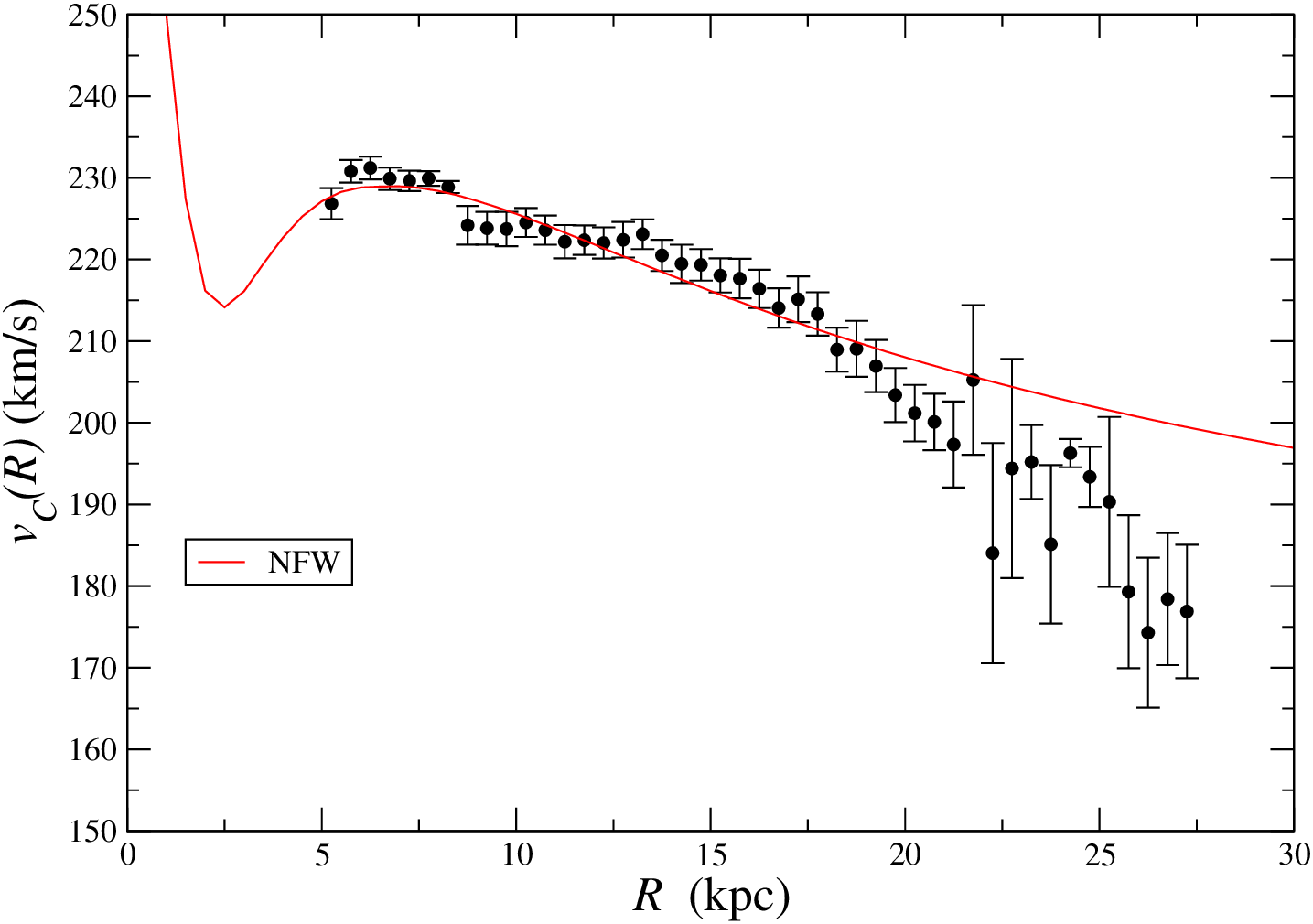}
\caption{
Best fit of the NFW mass model 
to the   rotation curve given in Tab.\ref{tabrot} (DR3+ determination of the rotation curve). 
}
 \label{NFW_fits} 
\end{figure}

The best fit model is found by minimizing the reduced $\chi^2_\nu$,    
where $\nu = N_{data} - 2$ as there  are two free parameters in the model.
Results  are shown in Fig.\ref{NFW_fits}: note that the increase of the rotation curve for small radii is due to the effect of the bulge. 
In Tab.\ref{tabnfw} we report the values  of the NFW best fit parameters.
When we use the  determination of the rotation curve
by \cite{Eilers_etal_2019} 
 our best fit parameters coincide, within the error bars, with theirs
(note that, in this case, 
we used $M_{bulge}  \approx 1 \times 10^{10} M_\odot$ to make 
a fit directly comparable).
The  values 
of $R_{vir}$ and $M_{vir}$ are the smallest 
for the DR3 determination of the rotation curve that does not  cover
 the range of radii for  $R<8$ kpc; 
 for the case DR3+, they are intermediate between the DR3 and the E19 case. 
{  The variations of the fitted parameters that we find in the different sample have a simple explanantion is terms of the 
finite range of  radii accessible to observation  that extend to only about 15\% of $R_{vir} \approx 200$ kpc and because 
the data for $R > 20$ kpc  have little leverage on the fit (see discussion in \cite{deBlok_etal_2001a}).}
Considering the stellar contributions, the total mass of the MW  (i.e., halo and stellar components) 
for the best fit of  DR3+ rotation curve 
 is thus $(73 \pm 3) \times 10^{10} M_\odot$
inside a virial radius of $R_{vir} = 183 \pm 1.84$ kpc.
\begin{table} 
\begin{center}
\begin{tabular}{ c c c c c c  }
\hline 
{  RC}  &  $M_{vir}$ & $R_{vir}$ & {  $c$ }                  & {  $c_{NFW}$} &  $\chi_\nu^2$    \\  
\hline 
 E19                       & $80 \pm 2$                             &  $197 \pm 2$    & $13.0\pm 0.5$     & $7.2\pm 1$       &  2.1      \\  
\hline 
 DR3                     & $48 \pm 3$                             & $166 \pm 3$     & $19.6 \pm 0.5$     & $7.6\pm 1$        & 2.4   \\  
\hline 
 DR3+                   & $65\pm 3$                               & $183 \pm 3$     &  $14.5\pm 0.5$    &  $7.3\pm 1 $        & 1.8   \\  
\hline 
\end{tabular}
\end{center}
\caption{  Results of the best fits of a NFW model 
to the three determination of the rotation curve (RC) we considered (E19 is by  \cite{Eilers_etal_2019},
DR3 is by \cite{Wang_etal_2023}  and DR3+ is  presented in Tab.\ref{tabrot}  --- see text).
We used $H_0 = 69$ km/s/Mpc as the value of the Hubble's constant. 
The mass $M_{vir}$  in units $10^{10} M_\odot$ 
and the radius $R_{vir}$  in kpc 
} 
\label{tabnfw}
\end{table} 
In this model 
the DM halo becomes the dominant dynamical contribution for $R>15$ kpc,
whereas the inner part is dominated
by the stellar components.

{  
Normally,  NFW profiles
are characterized by 2 model parameters \citep{Navarro_etal_1997}, and mass model 
fits usually consider both parameters as free (see, e.g., \cite{deBlok_etal_2008,Eilers_etal_2019}). 
However, it was then shown by \cite{Maccio_etal_2008} that the N-body calculations
which resulted in the NFW profile model clearly show that the concentration parameter $c$ is
not an independent parameter but is in fact strongly correlated
with $V_{vir}$, the rotational velocity at $R_{vir}$   can be written as 
\citep{Hessman+Ziebart_2011} 
\be
\label{cnfw} 
c_{NFW} \approx 7.80 
\left(
\frac{V_{vir}} {100 \; \; \mbox{km s}^{-1}}  
\right)^{-0.294}
\ee
(derived from  Eq. (10) of \cite{Maccio_etal_2008} --- see  \cite{Dutton_etal_2014} for a slightly different phenomenological fit). 
Including this intrinsic correlation reduces the number of fit parameters by one. In Tab.\ref{tabnfw}, the values of the best-fit concentration parameter $c$ and $c_{NFW}$ computed from Eq.\ref{cnfw} by inserting the value of the rotational velocity at $R_{vir}$ are reported. It is noticed that in all cases these are not consistent with the predictions of $\Lambda$CDM, which provides the well-defined mass-concentration relation Eq.\ref{cnfw}. We thus find that the Galactic halo has a concentration parameter $c \in (13,20)$, which is higher than theoretical expectations based on cosmological simulations \citep{Maccio_etal_2008}. High values of the concentration parameter have also been found by other studies \cite{Bovy_etal_2012,Deason_etal_2012,Kafle_etal_2014,McMillan_2017,Monari_etal_2018,Lin+Li_2019,Eilers_etal_2019} that are in tension with theoretical expectations based on cosmological simulations.

It should be noted that the potential systematic effect of adiabatic compression was not taken into account in the mass models. Adiabatic compression is an inevitable mechanism that occurs when a luminous galaxy is formed within a DM halo \citep{Gnedin_etal_2004,Sellwood+McGaugh_2005}. This process may help reconcile the parameters found with the Eq.\ref{cnfw} as it has the effect of raising the effective concentration of the resulting halo above that of the primordial initial condition to which the mass-concentration relation applies \citep{McGaugh_2016}. This means that the higher than expected concentration of the MW halo could be a manifestation of the contraction of the DM halo induced by the presence of a galaxy at its center \citep{Cautun_etal_2020}. However, it should be noted that there are contradictory statements about the adiabatic compression of our Galaxy's dark halo in the literature, for instance, \cite{Binney+Piffl_2015} finds evidence that rules it out.} 


\subsection{The Dark Matter Disk model} 

As discussed above the "Bosma 
effect"  assumes a correlation between 
the distribution of DM and that of the 
gas so that the rotation curve can be written as 
\be
\label{dmd0} 
v_c^2 = v_{thin}^2 +  v_{thick}^2+ v_{bulge}^2 + \Upsilon_{gas} v_{gas}^2 \;, 
\ee
where $v_{gas}$ is the circular velocity of the gas and $ \Upsilon_{gas}$, the ratio between
the DM and gas mass,  is an appropriate rescaling factor 
that must be determined in order to fit the observed rotation curve. 

For the case of external galaxies it is observed only the distribution of neutral H\RNum{1}, and not that of total gas;  
for this reason \cite{Hessman+Ziebart_2011} parametrized the model as 
\be
\label{dmd1} 
v_c^2 =\Upsilon^*(v_{thin}^2 +  v_{thick}^2+ v_{bulge}^2) + \Upsilon_{HI} v_{HI}^2 
\ee
where $\Upsilon^*$ is an additional free parameter  introduced because the  H\RNum{1} surface density 
does not reflect the total gas surface density in the inner galactic region
 and   the stellar disk was used as proxy 
of a dark mass component. 
In the case of the MW
there are available both the radial atomic (HI) gas    
\citep{Kalberla+Kerp_2009}, 
and the molecular gas (H$_2$) surface density profiles 
\citep{Bigiel+Blitz_2012}.

We consider two different functional behaviors for the gas surface density:
(i) An exponential surface density on a thin disk (TD) (ii) The sum of that of H\RNum{1} and H$_2$ confined on a TD.
In the first case, the free parameters are the exponential length scale $R_d$ and the total mass $M_d$, while in the second case, only the mass is considered. The comparison between these two different behaviors of the gas surface density allows us to understand the effect of the functional dependence of the gas component on the final mass estimate.

\begin{figure} 
\includegraphics[width=3.5in]{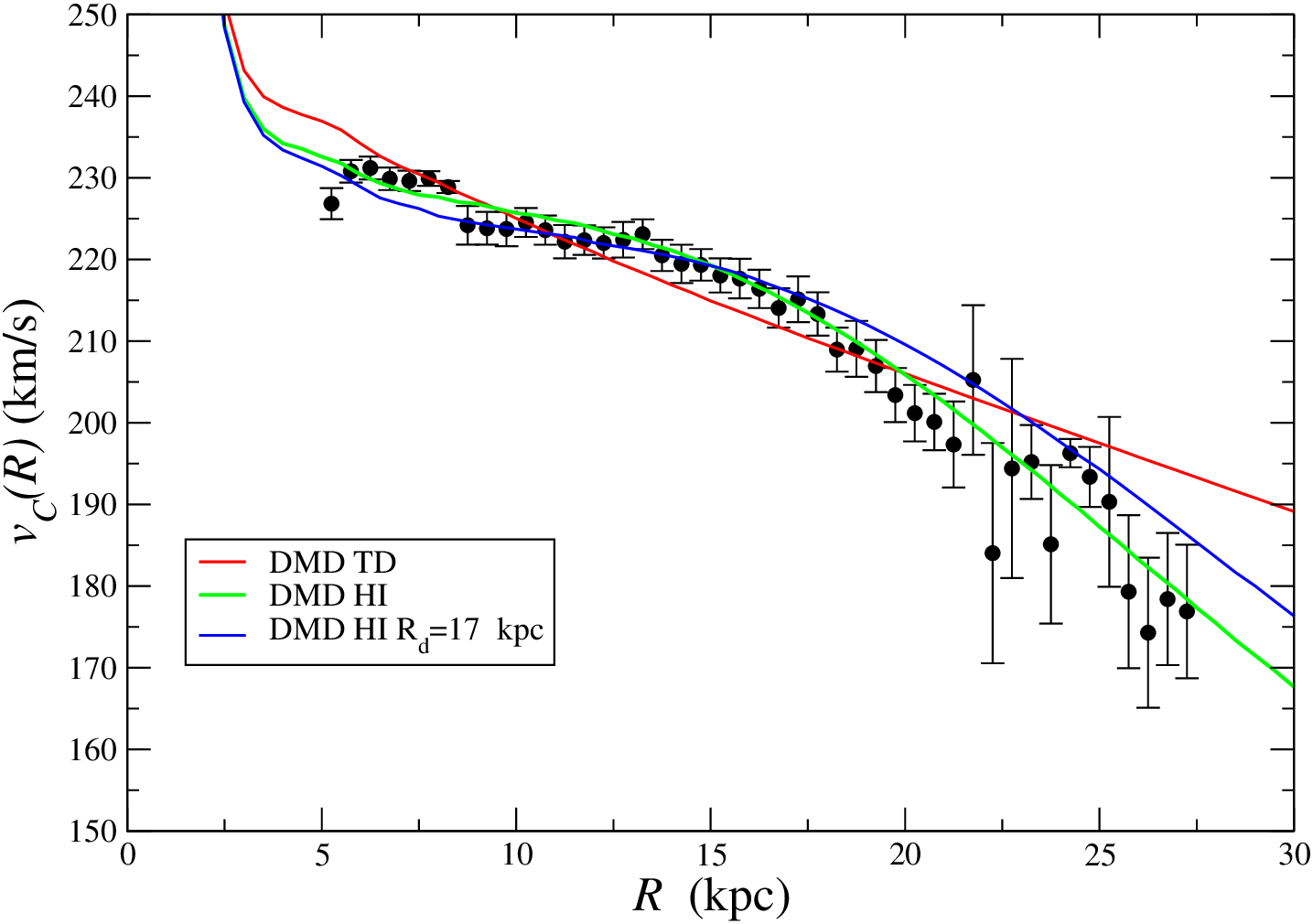}
\caption{
Best fit of the DMD  mass model  (i.e., Eq.\ref{dmd1}) 
to the   rotation curve given in Tab.\ref{tabrot} (DR3+ rotation curve):
we show results for an exponentially decaying surface mass 
on a thin disk (Eqs.\ref{sigma}-\ref{eq:td}),  
the case in which the surface density
is given by Eq.\ref{SigmaSHI}  with $R_d$ and $M_d$ as free parameters 
and the same but with $R_d=17$ kpc 
(corresponding to the value measured for the distribution of Galactic HI) 
and only the DM disk's mass $M_d$ as free parameter. 
}
 \label{fits-DMD} 
\end{figure}

For the first case, we used the well-known result 
 that 
the circular velocity generated 
by an exponentially decaying surface mass density 
\be
\label{sigma} 
\Sigma(R) = \Sigma_0 \exp\left( -\frac{R}{R_d} \right)\;,
\ee
(with total mass total equal to $M_d = 2 \pi \Sigma_0 R_d^2$) 
constrained on a TD, is  (see, e.g., \cite{Binney_Tremaine_2008})
\begin{equation}
\label{eq:td}
v_c^2(R)=4 \pi G \Sigma_0 R_d y^2[I_0(y)K_0(y)-I_1(y)K_1(y)],
\end{equation}
where $y=R/2R_d$ and $I_n$ and $K_n$ are the modified Bessel functions.
Results for  this model are shown in Fig.\ref{fits-DMD} and in Tab.\ref{tab:fits-DMD}:
the smallest  value of $\chi_\nu^2$ is  found for the DR3+ rotation curve.
This is 
smaller than for the NFW model: this can be seen by comparing the large radius behaviors 
in Figs.\ref{NFW_fits}-\ref{fits-DMD}. As in the previous case, 
the increase of the rotation curve for small radii is due to the effect of the bulge.
The estimated mass 
of the DMD, i.e., $M_{DMD} \approx 15 \times 10^{10} M_\odot$  
(with a characteristic scale-length $R_d \approx 15$ kpc) 
is about a factor 2  larger than the mass of all the stellar components (i.e., $M_{stellar} \approx 8\times 10^{10} M_\odot$)   and 
it is about  factor 7 smaller than the virial mass of the NFW halo mass model.
Note the Galaxy's mass in NFW model must include the contribution of the halo up to the 
virial radius, i.e., $R_{vir} \approx 180$ kpc; instead, by assuming that  DM  is confined on a disk
the mass of the galaxy $M_d$ is the one corresponding to a characteristic disk's radius $R_d \approx 15$ kpc. 

\begin{table} 
\begin{center}
\begin{tabular}{ c c c c }
\hline 
RC    & $R_d$       & $M_d$          & $\chi_\nu^2$\\  
\hline 
 E19        & $10.6 \pm 0.2$ &$16.3 \pm 1 $ & 2.0      \\  
\hline
 DR3      & $8.8 \pm 0.2$ &$12.4 \pm 1 $ & 1.4      \\  
 \hline
 DR3+     &  $10.2 \pm 0.2$ &$15.2 \pm 1$ & 1.3      \\  
\hline 
\end{tabular}
\end{center}
\caption{Results of the best fits of a DMD model  (i.e., Eq.\ref{dmd1}) 
for a exponentially decaying surface mass 
density on a thin disk (Eqs.\ref{sigma}-\ref{eq:td}) 
to the three determinations of the rotation curve we considered.
Units are $R_d$ in kpc  and $M_d$
in  $10^{10} M_\odot$. } 
\label{tab:fits-DMD}
\end{table} 

\begin{figure} 
\includegraphics[width=3.5in]{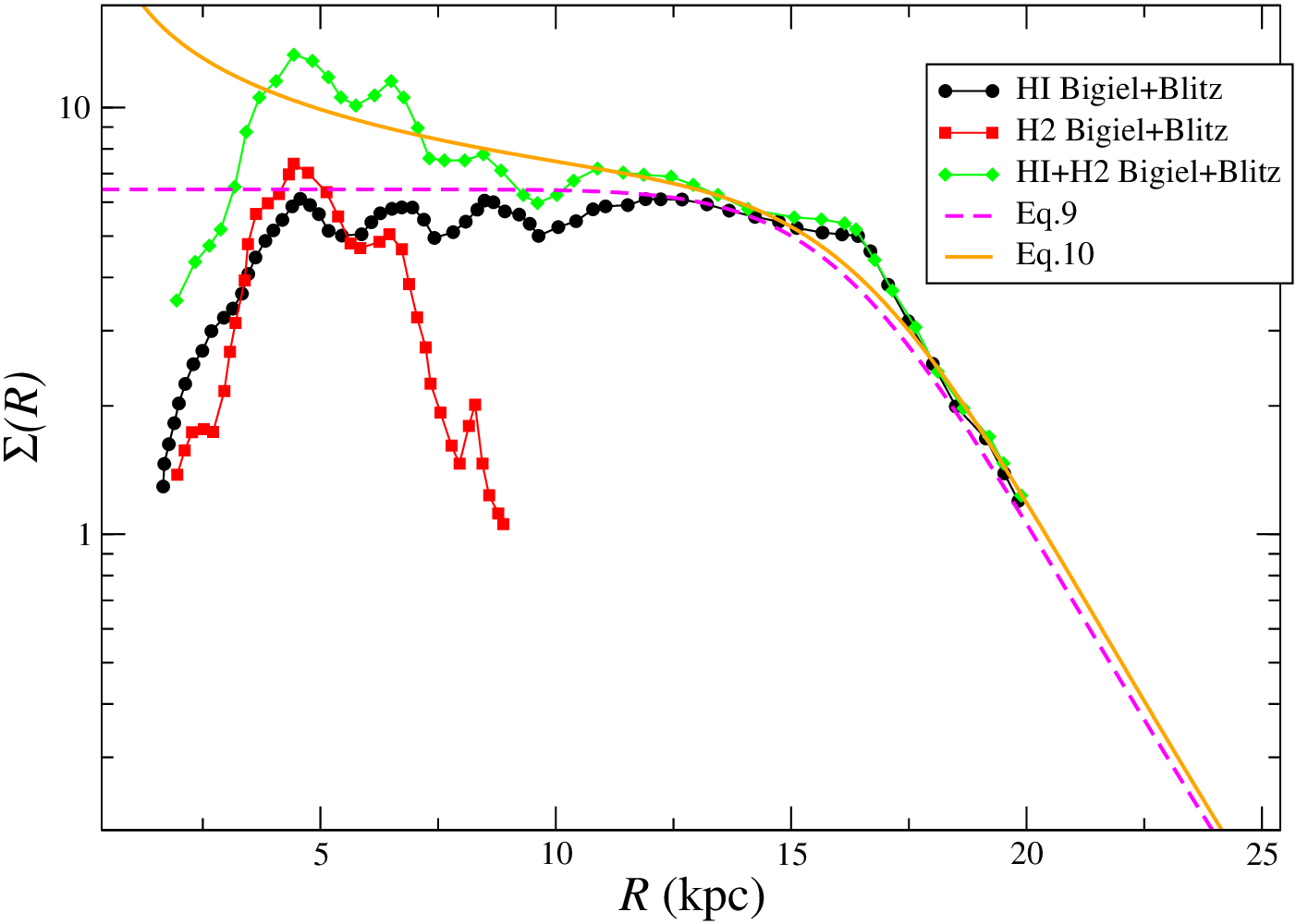}
\includegraphics[width=3.5in]{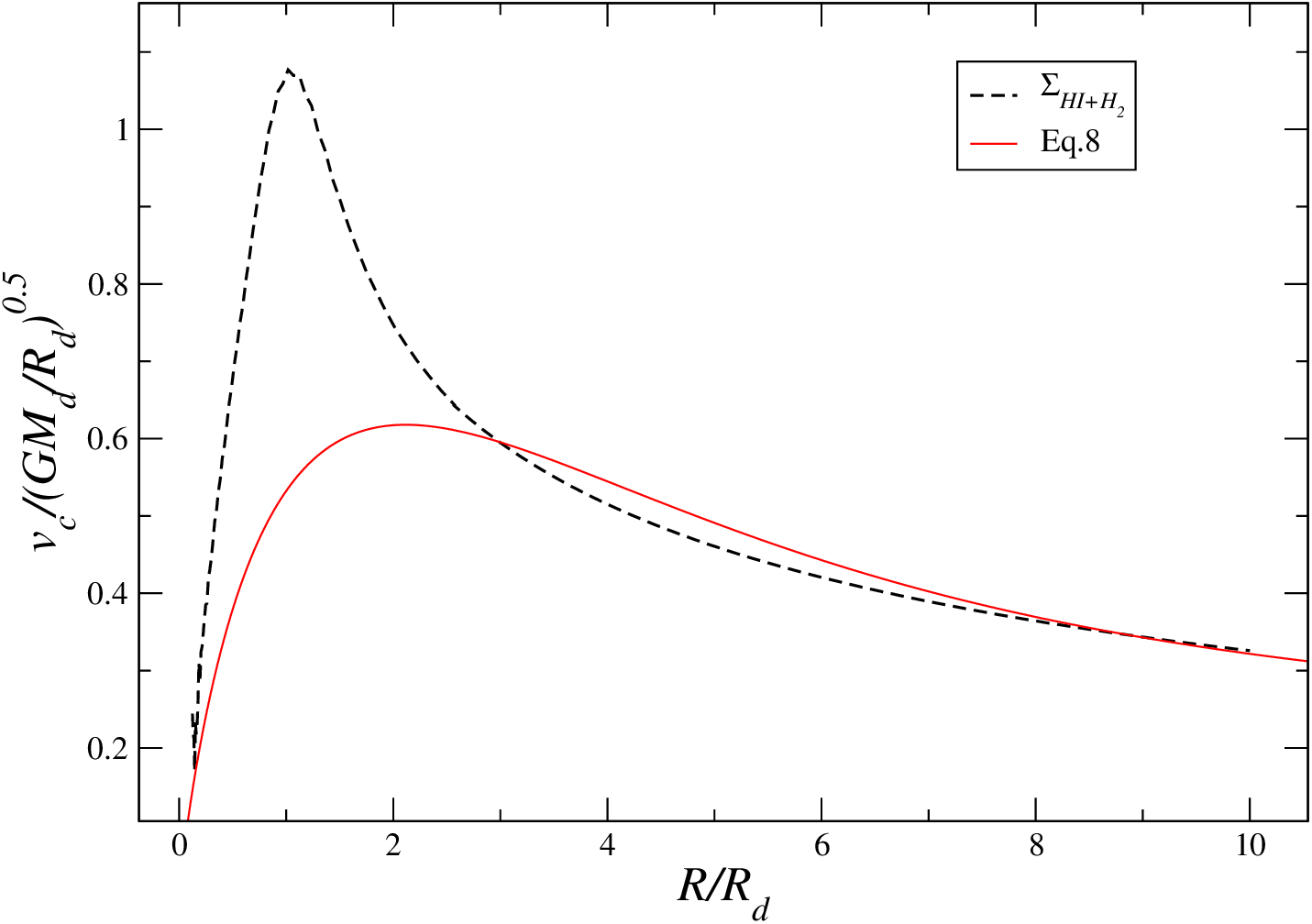}
\caption{
Upper panel: 
surface density profile  in  Eq.\ref{SigmaSHI} (solid line) (in $M_\odot $ pc$^{-2}$) and data 
 from \cite{Kalberla+Kerp_2009,Bigiel+Blitz_2012}.  
Bottom panel: circular velocity  of a distribution of matter confined on a thin disk 
with surface density profile in. Eq.\ref{SigmaSHI} 
together with the behavior of Eq.\ref{eq:td}.
}
 \label{SBHI} 
\end{figure}

We now assume that the observed atomic  H\RNum{1} and molecular H$_2$ distribution
traces that of DM. 
We found that a
 useful approximation to the observed surface density of HI, reported in 
 \cite{Kalberla+Kerp_2009,Bigiel+Blitz_2012}, 
is given by (see  the upper panel of Fig.\ref{SBHI}) 
\be
\label{SigmaSHI} 
\Sigma_{HI}(R) = \frac{\Sigma_0 }{1+\left( \frac{R}{R_d} \right)^\alpha}
\ee
with $\Sigma_0 \approx 6 \; M_\odot$ pc$^{-2}$ , $R_d=17$ kpc and $\alpha=10$
(this corresponds to a total  H\RNum{1} mass of $M_{HI} \approx 0.5 \times 10^{10} M_\odot$).
For the observed surface density of HI+H$_2$  \citep{Bigiel+Blitz_2012}, 
we find that there is small difference at small radii, i.e. $R<4$ kpc,  that we parametrize, for $R>2$ kpc,  as  
\be
\label{SigmaSHI+H2} 
\Sigma_{HI+H_2}(R) = \frac{\Sigma_0 } { \left(1+\left( \frac{R}{R_d} \right)^\alpha \right)  \left( \frac{R}{R_d} \right)^{0.25} }\;. 
\ee
Note that these approximations are useful only as they give an analytical reference but are not used in what follows
to compute the circular velocity.

Given the surface density profile in Eq.\ref{SigmaSHI+H2}, it is possible to compute the corresponding gravitational potential and, from it, the circular velocity $v_{c}$. Unlike the case of a spherical mass distribution, in the case of a disk, the result that the mass distribution outside a particular radius does not affect the effective force inside that radius is not universally valid. For this reason, we numerically computed the circular velocity of a distribution of matter confined on a disk, with a height equal to 1/20 of its radius, and with the observed gas surface density profile. The behavior of the circular velocity is shown in Fig.\ref{SBHI} (bottom panel) together with the behavior for an exponentially decaying surface mass density confined on a TD (i.e., Eq.\ref{eq:td}).

The results of this model are shown in Fig.\ref{fits-DMD} and in Tab.\ref{tab:fits-DMD-HI}. Note that the value of the mass is smaller than the previous case by $\approx 40\%$, that is, the DM component is about the same of the visible mass component, i.e. $M_{DMD} \approx 9 \times 10^{10} M_\odot$, and it is about a factor 6 smaller than the viral mass of the NFW halo. Finally, we note that the DM mass is about 20 times heavier than that of HI, as $M_{HI} \approx 0.5 \times 10^{10} M_\odot$: this value is comparable with what was found in external galaxies by \cite{Hessman+Ziebart_2011,Swaters_etal_2012}.

\begin{table} 
\begin{center}
\begin{tabular}{ c c c c c }
\hline 
RC     & $R_d$  & $M_d$  & $\chi_\nu^2$\\  
\hline 
 E19                & $11.6 \pm 0.2$       &$15.1 \pm 0.5$       & 2.6    \\  
\hline 
 DR3               & $15.6 \pm 0.5$     &$8.5 \pm 0.5$          & 1.2     \\  
\hline 
 DR3                & 17.0                     &  $10.0  \pm 0.5$     & 1.7      \\  
\hline 
 DR3+              & $15.5 \pm 0.5$    &$8.5 \pm 1 $            &   1.2     \\  
\hline 
 DR3+               & $17.0 $               &$9.7 \pm 1 $             & 2.3      \\  
\hline 
\end{tabular}
\end{center}
\caption{In this case the  surface density of dark matter is assumed to be traced by that of H\RNum{1}, i.e. it is proportional to 
Eq.\ref{SigmaSHI}. Two different cases are considered with the DR3 and DR3+ 
determinations of the rotation curve: in the first both $R_d$ (in kpc) and $M_d$
(in  $10^{10} M_\odot$)  are 
free parameters and in the second $R_d=17$ kpc, as measured from 
the distribution of Galactic H\RNum{1}, and $M_d$ is a free parameter.
 } 
\label{tab:fits-DMD-HI}
\end{table} 
%




\subsection{The MOND model}

{  Finally, we have fitted the rotation curve in the MOND framework \citep{Milgrom_1983}. It is worth noting that a similar study was presented in \cite{Chrobakova_etal_2020} using Gaia DR2 data, which reached distances up to $R \approx 20$ kpc and heights only up to $|Z| < 2$ kpc, where the decreasing trend in the MW rotating curve was less noticeable. Additionally, as discussed by \cite{Wang_etal_2023}, the binning of data in \cite{Chrobakova_etal_2020} was finer, leading to larger noise fluctuations and making the signal less reliable. Thus, the rotation curves of \cite{Chrobakova_etal_2020} are less robust than our current analysis, and the trends we see now were not noticed then.
}

\begin{figure} 
\includegraphics[width=3.5in]{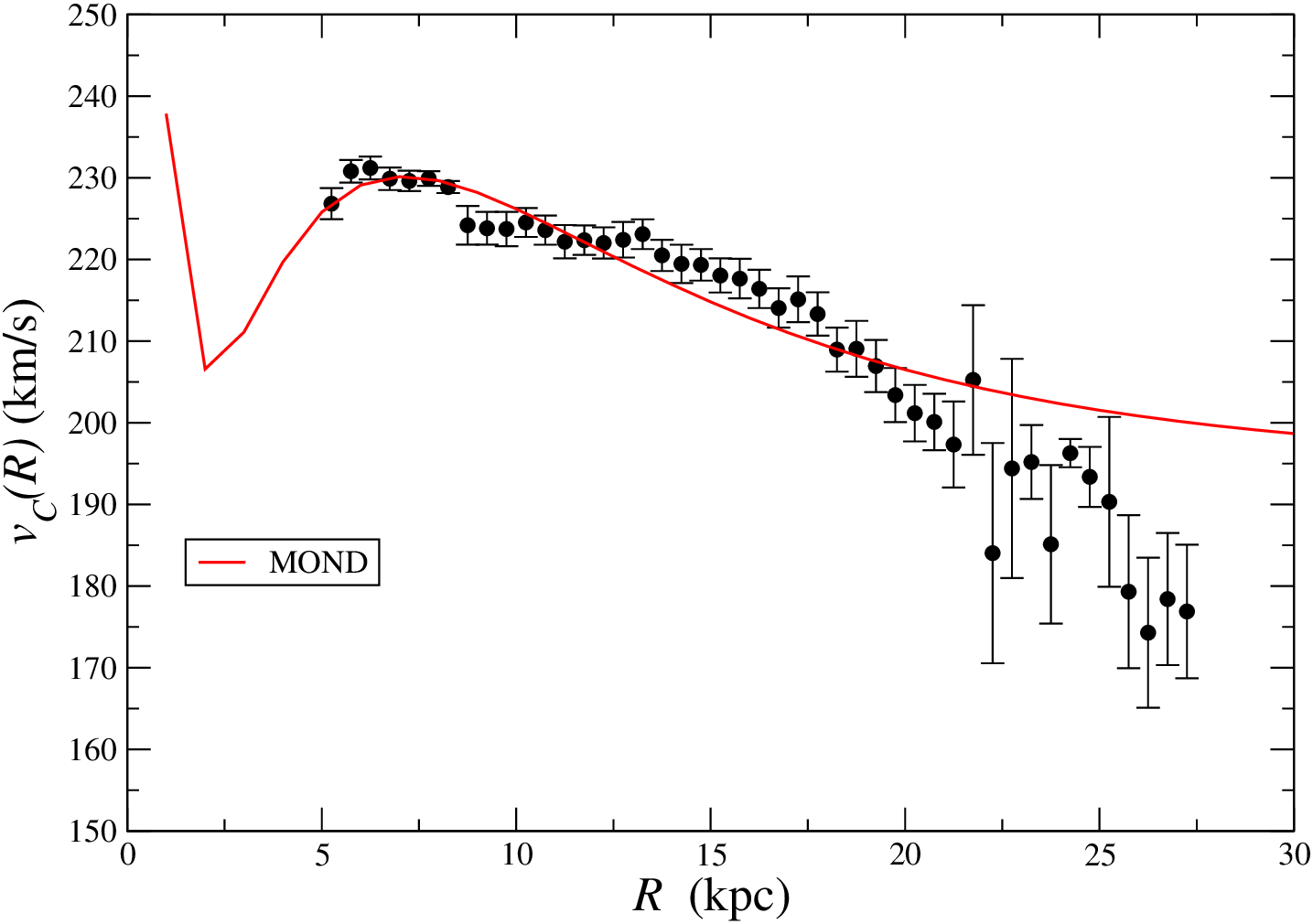}
\caption{Best fit of the MOND model to the rotation curve given in Table \ref{tabrot}. 
The red  curve represents MOND model with parameters fixed with values $R_{d}=3.1$ kpc, $M_{thin}=7.79\times 10^{10} M_\odot$.}
 \label{fig_mond} 
\end{figure}

We follow the approach of \cite{Chrobakova_etal_2020} and calculate the mondian acceleration as 
\begin{equation}
\label{mond_approx}
a_M=\sqrt{\frac{1}{2}a_N^2+\sqrt{\frac{1}{4}a_N^4+a_N^2a_0^2}}~,
\end{equation}
where the constant $a_0=1.2 \cdot 10^{-10}$ m  s$^{-2}$ \citep{scarpa06} 
the Newtonian acceleration $a_N$ can be calculated as \citep{McGaugh_etal2016}
\begin{equation}
a_N=\left|\frac{v_c^2(R)}{R}\right|~.
\end{equation}
{  where $v_c$ is the full, Newton-predicted rotation curve. Since this model does not include a halo, we consider $M_d$ and $R_d$ of the stellar disk as free parameters and we do not fit the bulge as this involves radii that are smaller than those sampled by the measured rotation curves. In Fig. \ref{fig_mond} we plot the best fit of the MOND model of the DR3+ determination of the rotation curve. 
The fit for the sample from the DR3 determination of the rotation curve is worse than the other two samples, as the DR3 sample lacks data for radii $R<10$, which are fitted better than the data at the largest radii.
In addition, the MOND model best fit has the minimal value of the $\chi_\nu^2$ only if we also use the disk's mass and radius as free parameters rather than using the fixed values of the stellar component as discussed above (see Section \ref{stellar}). We get that $M_d=(7.8 \pm 0.5) \times 10^{10} M_\odot$ and $R_d=(3.1 \pm 0.1)$ kpc: both values are larger than those used in Sect.\ref{stellar}.
}

However, note that Eq.\ref{mond_approx} is an approximation of the exact MOND relation, producing a small difference compared to the exact approach \citep{Lopez-Corredoira+Betancort-Rijo_2021}.  To estimate how much our solution deviates from the exact calculation, we use results of \cite{Lopez-Corredoira+Betancort-Rijo_2021}, 
who compare the difference between the approximation and the exact solution for an exponential disk (Figures 2 and 3 of their paper). 
Based on their result, we estimate that the rotation curve for our model deviates by $5-10\%$. 
Therefore, we make Monte Carlo simulations, taking into account this deviation and calculate new parameters of the fit, 
which we use to calculate the systematic errors of the free parameters (reported in Tab.\ref{tab:mond}).

\begin{table} 
\begin{center}
\begin{tabular}{ c c c c}
\hline 
 Sample     & $R_d$   & $M_d$  & $\chi_c^2$\\  
\hline 
 E19      &  \shortstack{$3.2 \pm 0.1$ (stat.) \\ $\pm 0.18$ (syst.)}    &   \shortstack{$7.97 \pm 0.1$ (stat.) \\ $\pm 0.9$ (syst.)}     &  1.46  \\  
\hline 
 DR3     & \shortstack{$2.8 \pm 0.29$ (stat.) \\ $\pm 0.2$ (syst.)}      & \shortstack{$7.24 \pm 1.5$ (stat.) \\ $\pm 1.0$ (syst.)}      & 5.79  \\  
\hline 
 DR3+    & \shortstack{$3.1 \pm 0.1$ (stat.) \\ $\pm 0.14$ (syst.)}      &   \shortstack{$7.79 \pm 0.5$ (stat.) \\ $\pm 0.7$ (syst.)}  &    2.32 \\ 
\hline 
\end{tabular}
\end{center}
\caption{Results of the best fit for the MOND model. $R_d$ is in kpc, $M_d$ is in units of $10^{10} M_\odot$. 
We report both the statistical (stat.) and systematic (syst.) errors. } 
\label{tab:mond}
\end{table} 

{  
Another aspect to consider is that the Jeans Equation used to derive the rotation curve in \cite{Wang_etal_2023}
assumes a Newtonian potential instead of MOND. Strictly one should use the Poisson equation 
\[ 
\nabla (\mu (x)\nabla \phi)=4\pi G\rho  \;, 
\]
where 
\[
x=\frac{|\vec{g}(\vec{r})|}{a_0}
\]
and $\vec{g}$ is the acceleration, 
instead of the Newtonian one $\nabla ^2\phi =4\pi G\rho $. 
However, the consequence of this MOND factor $\mu$ would be equivalent to adding a phantom mass to the real mass density distribution $\rho$ \citep{Lopez-Corredoira+Betancort-Rijo_2021}. The problem of the modification of Poisson equation implicitly included in Jeans equation is simply solved by changing the real density by an equivalent density including a phantom term (see Eq. 12 of \cite{Lopez-Corredoira+Betancort-Rijo_2021}). In our case, we have assumed an exponential function for $\rho$, and this is also an appropriate assumption for the equivalent density in MOND including the phantom term. Moreover, the outcome of the rotation speed is very little affected by the scalelength of the exponential distribution \citep{Chrobakova_etal_2020}. Therefore, we do not think the rotation speed might be significantly affected beyond the error bars due to MOND modification of Jeans equation.

In summary, the MW rotation curve is extremely well-fitted by MOND up to $R=19$ kpc. The region where MOND poorly fits the data is for $R>20$ kpc, i.e. where the rotation curve is found to decline both by \cite{Eilers_etal_2019} and \cite{Wang_etal_2023}. This result thus confirms earlier studies by \cite{McGaugh_2016,McGaugh_2019} at smaller radii, i.e. $R<10$ kpc. It is worth mentioning that \cite{McGaugh_2019} considered a model for the MW obtained by fitting the observed terminal velocities with the radial-acceleration relation. Such a model predicts a gradually declining rotation curve outside the solar circle with a slope of $-1.7$ km s$^{-1}$ kpc$^{-1}$, as subsequently observed by \cite{Eilers_etal_2019}.
  }




\section{Conclusions} 
\label{conclusions}

The Milky Way (MW) has several baryonic components, including a central nucleus, a bulge, and a disk. While many of their properties remain topics of debate, their masses are reasonably well-determined \citep{Bland-Hawthorn_etal_2016}. From kinematical and dynamical studies, we know that the mass of the MW must be larger than the sum of the baryonic components; indeed, to maintain the system in stable equilibrium with the observed amplitude of the circular velocity, a large fraction of the MW mass must be invisible, i.e., we cannot measure it directly but we can infer its presence by its gravitational influence.

Despite decades of intense efforts, the estimates of the mass of the MW still show significant scatter. These estimates are very sensitive to assumptions made in the modeling and, in particular, to the shape of the halo in which the Galaxy is embedded. Most mass estimators are limited to the region explored by the available tracer population, whose spatial distribution and kinematics are used to estimate the enclosed mass. Estimates of the MW's mass have been obtained based on the kinematics of halo stars, the kinematics of satellite galaxies and globular clusters, the evaluation of the local escape velocity, and the modeling of satellite galaxy tidal streams. Estimates typically range from as low as $0.5 \times 10^{12} M_\odot$ to as high as $4 \times 10^{12} M_\odot$ \citep{Bland-Hawthorn_etal_2016}. These estimates assume that dark matter (DM) is in a quasi-spherical virialized halo around the Galaxy \citep{Navarro_etal_1997,Sanders_2010}.

In this paper, we have presented a new estimation of the MW's virial mass in the framework of the NFW halo model. This estimation is based on combining two recent determinations of the Galaxy's rotation curve by \cite{Eilers_etal_2019}, in the range of 5-25 kpc, and by \cite{Wang_etal_2023}, in the range of 8-28 kpc, so as to obtain the rotation curve in the range of 5-28 kpc. In both cases, the Milky Way's rotation curve was measured in samples that are based, partially or completely, on data provided by the Gaia mission \citep{Gaia_2016}. These data have the unique characteristic of collecting the whole 6D spatial and velocity information of the sources with unprecedented precision and accuracy for the determination of their distances. Results for $v_c(R)$ by \cite{Eilers_etal_2019} and \cite{Wang_etal_2023} reasonably agree with each other and with another similar determination based on Gaia data but on a smaller distance range by \cite{Mroz_etal_2019}. In short, the Milky Way's rotation curve in these samples shows a gentle decline, passing from $\approx$ 230 km s$^{-1}$ at 5 kpc to $\approx$ 175 km s$^{-1}$ at 28 kpc.
{  The data with $R>20$ kpc are clearly important for the results of the fits, and one may wonder whether we are pushing the data to its limits. We think that this is not the case for two reasons. Firstly, the Lucy method, which is at the basis of the rotation curve determined by \cite{Wang_etal_2023}, has proven to be a solid technique that has given convergent results when passing from Gaia DR2 to Gaia DR3. The method works as long as errors in the parallax are Gaussian, and the fact that by lowering the errors, the results are convergent, means that this is not only a reasonable approximation but it is verified in the data. Secondly, \cite{Eilers_etal_2019} have already shown that there is a significant difference in the range 15-20 kpc with a flat rotation curve. Forthcoming data releases of the Gaia mission will provide more evidence that can possibly corroborate these results. } 

We find $M_{vir} = (6.5 \pm 3) \times 10^{11} M_\odot$ within a virial radius $R_{vir}=(180 \pm 3)$ kpc, which is $\approx 20\%$ smaller than the estimation by \cite{Eilers_etal_2019}. This gives a significantly lower mass estimation than what several previous studies suggest \citep{Bovy_etal_2012,Eadie+Harris_2016,Eadie_etal_2018}. This is due to the fact that the rotation curve measured by \cite{Eilers_etal_2019,Wang_etal_2023} showed a declining behavior up to 28 kpc, while most other determinations found $v_c(r) \approx$ const. at the same radii (see, e.g., \cite{Bhattacharjee_etal_2014,Sofue_2020} and references therein).

{  
We have then considered an additional phenomenological constraint derived from N-body calculations \citep{Maccio_etal_2008}, which reduces the number of free parameters of the NFW profile from two to one. We find that in this case the NFW fit parameters fall outside the predicted range of mass and concentration, so that with this additional constraint (see Eq.\ref{cnfw}), there is a tension with expectations from simulations.
}

We then considered an alternative model for the distribution of DM  in the Galaxy, called as the Dark Matter Disk (DMD) model. This model is inspired by the "Bosma Effect"   \citep{Bosma_1978,Bosma_1981} and assumes that the DM component is confined to a disk and is traced by the gas distribution. We find that the amount of DM in this model is a factor of 9 smaller than the NFW case, being about the same as the visible mass component. We also found that the DMD models are statistically as good as the NFW ones.
As DM is confined on a disk and not distributed in a spherical halo, 
it is not surprising that
we find that its amount is a factor 9 smaller than the NFW case 
being about the same of the visible mass component, i.e. 
$M_{stellar} \approx 8\times 10^{10} M_\odot \approx M_{DMD}$ so that 
the total mass of the Galaxy in this case is $1.6 \times 10^{11} M_\odot$. 
In this case the characteristic scale-length of the disk is $R_d=17$ kpc
and the DM mass is about 20 times heavier than that of HI, as 
$M_{HI} \approx 0.5 \times 10^{10} M_\odot$: this value 
is in line with what found in external galaxies by  \cite{Hessman+Ziebart_2011,Swaters_etal_2012}. 
{  The large radius behavior of the rotation curve in this model is determined by that of the rescaled gas component. For this reason it is worth noting that \cite{Bigiel+Blitz_2012} found that the azimuthally averaged radial distribution of the neutral gas surface density in 
a sample of  nearby spiral galaxies that includes the Milky Way, exhibits a well-constrained universal exponential distribution beyond $0.2 \times r_{25}$: in the framework of the DMD model, this universal gas profile  corresponds to the same large radius 
(in units of $ r_{25}$) rotation curve shape.}

We find that the DMD models are statistically as good as
the NFW ones. 
{  It should be emphasized that the DMD fits would not be as successful if the rotation curve did not show the decreasing behavior observed in samples based on the Gaia data-set. } 

{ 
Finally, we considered a model based on the Modified Newtonian Dynamics (MOND) framework
\citep{Milgrom_1983,scarpa06,McGaugh_etal2016}, which does not assume the presence of a heavy DM halo or a DM disk but instead hypothesizes a slower decay of the gravitational force to equilibrate the rotation velocity with the observed stellar mass components. We found that the data from the rotation curve of the MW is well-described by the MOND mass model up to a distance of 19 kpc
as previously fond by, e.g., \cite{McGaugh_etal2016}.
As for the case of the NFW model the data for $R>20$ kpc agree less well with the model because the decreasing behavior of the rotation curve.
}

Overall, we can conclude that, from the point of view of compatibility with the observations of the rotation curve, the DMD hypothesis represents a plausible galactic model. However, this model requires further studies, and it is important to investigate the nature of the matter that can be confined in the disk and whether any possible candidate is compatible with the present state of the art. Additionally, a refined modeling of this possible DM component and its compatibility with present observational constraints must be further investigated. In this respect, it is worth recalling that \cite{Pfenniger_etal_1994} suggested that such dark component, or a fraction of it, might be in molecular form and distributed in cold clouds that still elude direct detection. A refined modeling of this possible DM component and its compatibility with present observational constraints is certainly worth investigating.  %
{  

The near proportionality between HI and DM in outer galactic disks suggests that DM can be distributed on a disk rather than in a spherical halo. However, adding baryons in the disk of galaxies poses the problem of disk stability. It is well-known that self-gravitating disks close to stationary equilibrium and dominated by rotational motions are remarkably responsive to small disturbances
(see, e.g., 
\cite{Sellwood+Carlberg_1984,Sellwood+Carlberg_2014,Sellwood+Carlberg_2019,Binney_Tremaine_2008,Dobbs_etal_2018} 
and references therein). \cite{Revaz_etal_2009} have shown that the global stability is ensured if the interstellar medium is multi-phased, composed of two partially coupled phases: a visible warm gas phase and a weakly collisionless cold dark phase corresponding to a fraction of the unseen baryons. This new model still possesses a DM halo as in CDM ones.
A different theoretical scenario occurs if the disk is originated by a top-down gravitational collapse of an isolated over-density.  \cite{Benhaiem+SylosLabini+Joyce_2019,SylosLabini_etal_2020}  have shown that this scenario may improve the resistance to the effect of internal or external perturbations. A more detailed investigation of the stability of these systems and of a cosmological model for their formation will be presented in a forthcoming work.
}

\section*{Acknowledgments}

FSL thanks Frederic Hessman for many useful 
comments and suggestions. We thank 
S\'ebastien Comer\'on, 
Daniel Pfenniger and Hai-Feng Wang 
for useful discussions.  
{  We also thank an anonymous referee for a number of comments 
and suggestions which have allowed us to improve the presentation of our results.} 
This work has made use of data from the European Space Agency (ESA) mission
{\it Gaia} (\url{https://www.cosmos.esa.int/gaia}), processed by the {\it Gaia}
Data Processing and Analysis Consortium (DPAC,
\url{https://www.cosmos.esa.int/web/gaia/dpac/consortium}). Funding for the DPAC
has been provided by national institutions, in particular the institutions
participating in the {\it Gaia} Multilateral Agreement.


\begin{thebibliography}{}
\expandafter\ifx\csname natexlab\endcsname\relax\def\natexlab#1{#1}\fi
\providecommand{\url}[1]{\href{#1}{#1}}
\providecommand{\dodoi}[1]{doi:~\href{http://doi.org/#1}{\nolinkurl{#1}}}
\providecommand{\doeprint}[1]{\href{http://ascl.net/#1}{\nolinkurl{http://ascl.net/#1}}}
\providecommand{\doarXiv}[1]{\href{https://arxiv.org/abs/#1}{\nolinkurl{https://arxiv.org/abs/#1}}}

\bibitem[{{Antoja} {et~al.}(2018){Antoja}, {Helmi}, {Romero-G{\'o}mez}, {Katz},
  {Babusiaux}, {Drimmel}, {Evans}, {Figueras}, {Poggio}, {Reyl{\'e}}, {Robin},
  {Seabroke}, \& {Soubiran}}]{Antoja_etal_2018}
{Antoja}, T., {Helmi}, A., {Romero-G{\'o}mez}, M., {et~al.} 2018, Nature, 561,
  360, \dodoi{10.1038/s41586-018-0510-7}

\bibitem[{{Benhaiem} {et~al.}(2019){Benhaiem}, {Sylos Labini}, \&
  {Joyce}}]{Benhaiem+SylosLabini+Joyce_2019}
{Benhaiem}, D., {Sylos Labini}, F., \& {Joyce}, M. 2019, Phys.Rev.E, 99,
  022125, \dodoi{10.1103/PhysRevE.99.022125}

\bibitem[{{Bhattacharjee} {et~al.}(2014){Bhattacharjee}, {Chaudhury}, \&
  {Kundu}}]{Bhattacharjee_etal_2014}
{Bhattacharjee}, P., {Chaudhury}, S., \& {Kundu}, S. 2014, Astrophys.J., 785,
  63, \dodoi{10.1088/0004-637X/785/1/63}

\bibitem[{{Bigiel} \& {Blitz}(2012)}]{Bigiel+Blitz_2012}
{Bigiel}, F., \& {Blitz}, L. 2012, Astrophys.J., 756, 183,
  \dodoi{10.1088/0004-637X/756/2/183}

\bibitem[{{Binney} \& {Piffl}(2015)}]{Binney+Piffl_2015}
{Binney}, J., \& {Piffl}, T. 2015, Mon.Not.R.Astr.Soc., 454, 3653,
  \dodoi{10.1093/mnras/stv2225}

\bibitem[{Binney \& Tremaine(2008)}]{Binney_Tremaine_2008}
Binney, J., \& Tremaine, S. 2008, Galactic Dynamics (Princeton University
  Press)

\bibitem[{{Bland-Hawthorn} \& {Gerhard}(2016)}]{Bland-Hawthorn_etal_2016}
{Bland-Hawthorn}, J., \& {Gerhard}, O. 2016, Ann.Rev.Astron.Astrophys., 54,
  529, \dodoi{10.1146/annurev-astro-081915-023441}

\bibitem[{{Bosma}(1978)}]{Bosma_1978}
{Bosma}, A. 1978, PhD thesis, University of Groningen, Netherlands

\bibitem[{{Bosma}(1981)}]{Bosma_1981}
---. 1981, Astron.J., 86, 1791, \dodoi{10.1086/113062}

\bibitem[{{Bovy} {et~al.}(2012){Bovy}, {Allende Prieto}, {Beers}, {Bizyaev},
  {da Costa}, {Cunha}, {Ebelke}, {Eisenstein}, {Frinchaboy}, {Garc{\'\i}a
  P{\'e}rez}, {Girardi}, {Hearty}, {Hogg}, {Holtzman}, {Maia}, {Majewski},
  {Malanushenko}, {Malanushenko}, {M{\'e}sz{\'a}ros}, {Nidever}, {O'Connell},
  {O'Donnell}, {Oravetz}, {Pan}, {Rocha-Pinto}, {Schiavon}, {Schneider},
  {Schultheis}, {Skrutskie}, {Smith}, {Weinberg}, {Wilson}, \&
  {Zasowski}}]{Bovy_etal_2012}
{Bovy}, J., {Allende Prieto}, C., {Beers}, T.~C., {et~al.} 2012, Astrophys.J.,
  759, 131, \dodoi{10.1088/0004-637X/759/2/131}

\bibitem[{{Cautun} {et~al.}(2020){Cautun}, {Ben{\'\i}tez-Llambay}, {Deason},
  {Frenk}, {Fattahi}, {G{\'o}mez}, {Grand}, {Oman}, {Navarro}, \&
  {Simpson}}]{Cautun_etal_2020}
{Cautun}, M., {Ben{\'\i}tez-Llambay}, A., {Deason}, A.~J., {et~al.} 2020,
  Mon.Not.R.Astr.Soc., 494, 4291, \dodoi{10.1093/mnras/staa1017}

\bibitem[{{Chrob{\'a}kov{\'a}} {et~al.}(2020){Chrob{\'a}kov{\'a}},
  {L{\'o}pez-Corredoira}, {Sylos Labini}, {Wang}, \&
  {Nagy}}]{Chrobakova_etal_2020}
{Chrob{\'a}kov{\'a}}, {\v{Z}}., {L{\'o}pez-Corredoira}, M., {Sylos Labini}, F.,
  {Wang}, H.~F., \& {Nagy}, R. 2020, Astron.Astrohys, 642, A95,
  \dodoi{10.1051/0004-6361/202038736}

\bibitem[{{Chrob{\'a}kov{\'a}} {et~al.}(2022){Chrob{\'a}kov{\'a}}, {Nagy}, \&
  {L{\'o}pez-Corredoira}}]{Chrobakova_etal_2022}
{Chrob{\'a}kov{\'a}}, {\v{Z}}., {Nagy}, R., \& {L{\'o}pez-Corredoira}, M. 2022,
  Astron.Astrophys., 664, A58, \dodoi{10.1051/0004-6361/202243296}

\bibitem[{{de Blok} {et~al.}(2001){de Blok}, {McGaugh}, {Bosma}, \&
  {Rubin}}]{deBlok_etal_2001a}
{de Blok}, W.~J.~G., {McGaugh}, S.~S., {Bosma}, A., \& {Rubin}, V.~C. 2001,
  Astrophys.J.Lett., 552, L23, \dodoi{10.1086/320262}

\bibitem[{{De Blok} {et~al.}(2008){De Blok}, {Walter}, {Brinks},
  {Trachternach}, {Oh}, \& {Kennicutt}}]{deBlok_etal_2008}
{De Blok}, W.~J.~G., {Walter}, F., {Brinks}, E., {et~al.} 2008, Astronom.J.,
  136, 2648, \dodoi{10.1088/0004-6256/136/6/2648}

\bibitem[{{Deason} {et~al.}(2012){Deason}, {Belokurov}, {Evans}, \&
  {An}}]{Deason_etal_2012}
{Deason}, A.~J., {Belokurov}, V., {Evans}, N.~W., \& {An}, J. 2012,
  Mon.Not.R.Astr.Soc., 424, L44, \dodoi{10.1111/j.1745-3933.2012.01283.x}

\bibitem[{{Dobbs} {et~al.}(2018){Dobbs}, {Pettitt}, {Corbelli}, \&
  {Pringle}}]{Dobbs_etal_2018}
{Dobbs}, C.~L., {Pettitt}, A.~R., {Corbelli}, E., \& {Pringle}, J.~E. 2018,
  Mon.Not.R.Astron.Soc., 478, 3793, \dodoi{10.1093/mnras/sty1231}

\bibitem[{{Dutton} \& {Macci{\`o}}(2014)}]{Dutton_etal_2014}
{Dutton}, A.~A., \& {Macci{\`o}}, A.~V. 2014, Mon.Not.R.Astron.Soc., 441, 3359,
  \dodoi{10.1093/mnras/stu742}

\bibitem[{{Eadie} {et~al.}(2018){Eadie}, {Keller}, \&
  {Harris}}]{Eadie_etal_2018}
{Eadie}, G., {Keller}, B., \& {Harris}, W.~E. 2018, Astrophys.J., 865, 72,
  \dodoi{10.3847/1538-4357/aadb95}

\bibitem[{{Eadie} \& {Harris}(2016)}]{Eadie+Harris_2016}
{Eadie}, G.~M., \& {Harris}, W.~E. 2016, Astrophys.J., 829, 108,
  \dodoi{10.3847/0004-637X/829/2/108}

\bibitem[{{Eilers} {et~al.}(2019){Eilers}, {Hogg}, {Rix}, \&
  {Ness}}]{Eilers_etal_2019}
{Eilers}, A.-C., {Hogg}, D.~W., {Rix}, H.-W., \& {Ness}, M.~K. 2019,
  Astrophys.J., 871, 120, \dodoi{10.3847/1538-4357/aaf648}

\bibitem[{{Gaia Collaboration} {et~al.}(2016){Gaia Collaboration}, {Prusti},
  {de Bruijne}, {Brown}, {Vallenari}, {Babusiaux}, {Bailer-Jones}, {Bastian},
  {Biermann}, {Evans}, \& et~al.}]{Gaia_2016}
{Gaia Collaboration}, {Prusti}, T., {de Bruijne}, J.~H.~J., {et~al.} 2016,
  Astron.Astrophys., 595, A1, \dodoi{10.1051/0004-6361/201629272}

\bibitem[{{Gaia Collaboration} {et~al.}(2018){Gaia Collaboration}, {Katz},
  {Antoja}, {Romero-G{\'o}mez}, {Drimmel}, {Reyl{\'e}}, {Seabroke}, {Soubiran},
  {Babusiaux}, {Di Matteo}, \& et~al.}]{Katz_etal_2018}
{Gaia Collaboration}, {Katz}, D., {Antoja}, T., {et~al.} 2018,
  Astron.Astrophys., 616, A11, \dodoi{10.1051/0004-6361/201832865}

\bibitem[{{Gaia Collaboration} {et~al.}(2022){Gaia Collaboration}, {Vallenari},
  {Brown}, {Prusti}, {de Bruijne}, {Arenou}, {Babusiaux}, {Biermann},
  {Creevey}, {Ducourant}, {Evans}, {Eyer}, {Guerra}, {Hutton}, {Jordi},
  {Klioner}, {Lammers}, {Lindegren}, {Luri}, {Mignard}, {Panem}, {Pourbaix},
  {Randich}, {Sartoretti}, {Soubiran}, {Tanga}, {Walton}, {Bailer-Jones},
  {Bastian}, {Drimmel}, {Jansen}, {Katz}, {Lattanzi}, {van Leeuwen}, {Bakker},
  {Cacciari}, {Casta{\~n}eda}, {De Angeli}, {Fabricius}, {Fouesneau},
  {Fr{\'e}mat}, {Galluccio}, {Guerrier}, {Heiter}, {Masana}, {Messineo},
  {Mowlavi}, {Nicolas}, {Nienartowicz}, {Pailler}, {Panuzzo}, {Riclet}, {Roux},
  {Seabroke}, {Sordo{\o}rcit}, {Th{\'e}venin}, {Gracia-Abril}, {Portell},
  {Teyssier}, {Altmann}, {Andrae}, {Audard}, {Bellas-Velidis}, {Benson},
  {Berthier}, {Blomme}, {Burgess}, {Busonero}, {Busso}, {C{\'a}novas}, {Carry},
  {Cellino}, {Cheek}, {Clementini}, {Damerdji}, {Davidson}, {de Teodoro},
  {Nu{\~n}ez Campos}, {Delchambre}, {Dell'Oro}, {Esquej},
  {Fern{\'a}ndez-Hern{\'a}ndez}, {Fraile}, {Garabato}, {Garc{\'\i}a-Lario},
  {Gosset}, {Haigron}, {Halbwachs}, {Hambly}, {Harrison}, {Hern{\'a}ndez},
  {Hestroffer}, {Hodgkin}, {Holl}, {Jan{\ss}en}, {Jevardat de Fombelle},
  {Jordan}, {Krone-Martins}, {Lanzafame}, {L{\"o}ffler}, {Marchal}, {Marrese},
  {Moitinho}, {Muinonen}, {Osborne}, {Pancino}, {Pauwels}, {Recio-Blanco},
  {Reyl{\'e}}, {Riello}, {Rimoldini}, {Roegiers}, {Rybizki}, {Sarro}, {Siopis},
  {Smith}, {Sozzetti}, {Utrilla}, {van Leeuwen}, {Abbas}, {{\'A}brah{\'a}m},
  {Abreu Aramburu}, {Aerts}, {Aguado}, {Ajaj}, {Aldea-Montero}, {Altavilla},
  {{\'A}lvarez}, {Alves}, {Anders}, {Anderson}, {Anglada Varela}, {Antoja},
  {Baines}, {Baker}, {Balaguer-N{\'u}{\~n}ez}, {Balbinot}, {Balog}, {Barache},
  {Barbato}, {Barros}, {Barstow}, {Bartolom{\'e}}, {Bassilana}, {Bauchet},
  {Becciani}, {Bellazzini}, {Berihuete}, {Bernet}, {Bertone}, {Bianchi},
  {Binnenfeld}, {Blanco-Cuaresma}, {Blazere}, {Boch}, {Bombrun}, {Bossini},
  {Bouquillon}, {Bragaglia}, {Bramante}, {Breedt}, {Bressan}, {Brouillet},
  {Brugaletta}, {Bucciarelli}, {Burlacu}, {Butkevich}, {Buzzi}, {Caffau},
  {Cancelliere}, {Cantat-Gaudin}, {Carballo}, {Carlucci}, {Carnerero},
  {Carrasco}, {Casamiquela}, {Castellani}, {Castro-Ginard}, {Chaoul},
  {Charlot}, {Chemin}, {Chiaramida}, {Chiavassa}, {Chornay}, {Comoretto},
  {Contursi}, {Cooper}, {Cornez}, {Cowell}, {Crifo}, {Cropper}, {Crosta},
  {Crowley}, {Dafonte}, {Dapergolas}, {David}, {David}, {de Laverny}, {De
  Luise}, {De March}, {De Ridder}, {de Souza}, {de Torres}, {del Peloso}, {del
  Pozo}, {Delbo}, {Delgado}, {Delisle}, {Demouchy}, {Dharmawardena}, {Di
  Matteo}, {Diakite}, {Diener}, {Distefano}, {Dolding}, {Edvardsson}, {Enke},
  {Fabre}, {Fabrizio}, {Faigler}, {Fedorets}, {Fernique}, {Fienga}, {Figueras},
  {Fournier}, {Fouron}, {Fragkoudi}, {Gai}, {Garcia-Gutierrez},
  {Garcia-Reinaldos}, {Garc{\'\i}a-Torres}, {Garofalo}, {Gavel}, {Gavras},
  {Gerlach}, {Geyer}, {Giacobbe}, {Gilmore}, {Girona}, {Giuffrida}, {Gomel},
  {Gomez}, {Gonz{\'a}lez-N{\'u}{\~n}ez}, {Gonz{\'a}lez-Santamar{\'\i}a},
  {Gonz{\'a}lez-Vidal}, {Granvik}, {Guillout}, {Guiraud},
  {Guti{\'e}rrez-S{\'a}nchez}, {Guy}, {Hatzidimitriou}, {Hauser}, {Haywood},
  {Helmer}, {Helmi}, {Sarmiento}, {Hidalgo}, {Hilger}, {H{\l}adczuk}, {Hobbs},
  {Holland}, {Huckle}, {Jardine}, {Jasniewicz}, {Jean-Antoine Piccolo},
  {Jim{\'e}nez-Arranz}, {Jorissen}, {Juaristi Campillo}, {Julbe}, {Karbevska},
  {Kervella}, {Khanna}, {Kontizas}, {Kordopatis}, {Korn}, {K{\'o}sp{\'a}l},
  {Kostrzewa-Rutkowska}, {Kruszy{\'n}ska}, {Kun}, {Laizeau}, {Lambert},
  {Lanza}, {Lasne}, {Le Campion}, {Lebreton}, {Lebzelter}, {Leccia}, {Leclerc},
  {Lecoeur-Taibi}, {Liao}, {Licata}, {Lindstr{\o}m}, {Lister}, {Livanou},
  {Lobel}, {Lorca}, {Loup}, {Madrero Pardo}, {Magdaleno Romeo}, {Managau},
  {Mann}, {Manteiga}, {Marchant}, {Marconi}, {Marcos}, {Marcos Santos},
  {Mar{\'\i}n Pina}, {Marinoni}, {Marocco}, {Marshall}, {Polo},
  {Mart{\'\i}n-Fleitas}, {Marton}, {Mary}, {Masip}, {Massari},
  {Mastrobuono-Battisti}, {Mazeh}, {McMillan}, {Messina}, {Michalik}, {Millar},
  {Mints}, {Molina}, {Molinaro}, {Moln{\'a}r}, {Monari}, {Mongui{\'o}},
  {Montegriffo}, {Montero}, {Mor}, {Mora}, {Morbidelli}, {Morel}, {Morris},
  {Muraveva}, {Murphy}, {Musella}, {Nagy}, {Noval}, {Oca{\~n}a}, {Ogden},
  {Ordenovic}, {Osinde}, {Pagani}, {Pagano}, {Palaversa}, {Palicio},
  {Pallas-Quintela}, {Panahi}, {Payne-Wardenaar}, {Pe{\~n}alosa Esteller},
  {Penttil{\"a}}, {Pichon}, {Piersimoni}, {Pineau}, {Plachy}, {Plum}, {Poggio},
  {Pr{\v{s}}a}, {Pulone}, {Racero}, {Ragaini}, {Rainer}, {Raiteri}, {Rambaux},
  {Ramos}, {Ramos-Lerate}, {Re Fiorentin}, {Regibo}, {Richards}, {Rios Diaz},
  {Ripepi}, {Riva}, {Rix}, {Rixon}, {Robichon}, {Robin}, {Robin}, {Roelens},
  {Rogues}, {Rohrbasser}, {Romero-G{\'o}mez}, {Rowell}, {Royer}, {Ruz Mieres},
  {Rybicki}, {Sadowski}, {S{\'a}ez N{\'u}{\~n}ez}, {Sagrist{\`a} Sell{\'e}s},
  {Sahlmann}, {Salguero}, {Samaras}, {Sanchez Gimenez}, {Sanna},
  {Santove{\~n}a}, {Sarasso}, {Schultheis}, {Sciacca}, {Segol}, {Segovia},
  {S{\'e}gransan}, {Semeux}, {Shahaf}, {Siddiqui}, {Siebert}, {Siltala},
  {Silvelo}, {Slezak}, {Slezak}, {Smart}, {Snaith}, {Solano}, {Solitro},
  {Souami}, {Souchay}, {Spagna}, {Spina}, {Spoto}, {Steele},
  {Steidelm{\"u}ller}, {Stephenson}, {S{\"u}veges}, {Surdej}, {Szabados},
  {Szegedi-Elek}, {Taris}, {Taylo}, {Teixeira}, {Tolomei}, {Tonello}, {Torra},
  {Torra}, {Torralba Elipe}, {Trabucchi}, {Tsounis}, {Turon}, {Ulla}, {Unger},
  {Vaillant}, {van Dillen}, {van Reeven}, {Vanel}, {Vecchiato}, {Viala},
  {Vicente}, {Voutsinas}, {Weiler}, {Wevers}, {Wyrzykowski}, {Yoldas}, {Yvard},
  {Zhao}, {Zorec}, {Zucker}, \& {Zwitter}}]{Gaia_DR3}
{Gaia Collaboration}, {Vallenari}, A., {Brown}, A.~G.~A., {et~al.} 2022, arXiv
  e-prints, arXiv:2208.00211.
\newblock \doarXiv{2208.00211}

\bibitem[{{Gnedin} {et~al.}(2004){Gnedin}, {Kravtsov}, {Klypin}, \&
  {Nagai}}]{Gnedin_etal_2004}
{Gnedin}, O.~Y., {Kravtsov}, A.~V., {Klypin}, A.~A., \& {Nagai}, D. 2004,
  Astrophys.J., 616, 16, \dodoi{10.1086/424914}

\bibitem[{{GRAVITY Collaboration} {et~al.}(2018){GRAVITY Collaboration},
  {Abuter}, {Amorim}, {Anugu}, {Baub{\"o}ck}, {Benisty}, {Berger}, {Blind},
  {Bonnet}, {Brandner}, {Buron}, {Collin}, {Chapron}, {Cl{\'e}net}, {Coud{\'e}
  Du Foresto}, {de Zeeuw}, {Deen}, {Delplancke-Str{\"o}bele}, {Dembet},
  {Dexter}, {Duvert}, {Eckart}, {Eisenhauer}, {Finger}, {F{\"o}rster
  Schreiber}, {F{\'e}dou}, {Garcia}, {Garcia Lopez}, {Gao}, {Gendron},
  {Genzel}, {Gillessen}, {Gordo}, {Habibi}, {Haubois}, {Haug}, {Hau{\ss}mann},
  {Henning}, {Hippler}, {Horrobin}, {Hubert}, {Hubin}, {Jimenez Rosales},
  {Jochum}, {Jocou}, {Kaufer}, {Kellner}, {Kendrew}, {Kervella}, {Kok},
  {Kulas}, {Lacour}, {Lapeyr{\`e}re}, {Lazareff}, {Le Bouquin}, {L{\'e}na},
  {Lippa}, {Lenzen}, {M{\'e}rand}, {M{\"u}ler}, {Neumann}, {Ott}, {Palanca},
  {Paumard}, {Pasquini}, {Perraut}, {Perrin}, {Pfuhl}, {Plewa}, {Rabien},
  {Ram{\'\i}rez}, {Ramos}, {Rau}, {Rodr{\'\i}guez-Coira}, {Rohloff}, {Rousset},
  {Sanchez-Bermudez}, {Scheithauer}, {Sch{\"o}ller}, {Schuler}, {Spyromilio},
  {Straub}, {Straubmeier}, {Sturm}, {Tacconi}, {Tristram}, {Vincent}, {von
  Fellenberg}, {Wank}, {Waisberg}, {Widmann}, {Wieprecht}, {Wiest},
  {Wiezorrek}, {Woillez}, {Yazici}, {Ziegler}, \&
  {Zins}}]{Gravity_collaboration_2018}
{GRAVITY Collaboration}, {Abuter}, R., {Amorim}, A., {et~al.} 2018,
  Astron.Astrophys., 615, L15, \dodoi{10.1051/0004-6361/201833718}

\bibitem[{{Hessman} \& {Ziebart}(2011)}]{Hessman+Ziebart_2011}
{Hessman}, F.~V., \& {Ziebart}, M. 2011, Astron.Astrophys., 532, A121,
  \dodoi{10.1051/0004-6361/201117199}

\bibitem[{{Hoekstra} {et~al.}(2001){Hoekstra}, {van Albada}, \&
  {Sancisi}}]{Hoekstra_etal_2001}
{Hoekstra}, H., {van Albada}, T.~S., \& {Sancisi}, R. 2001,
  Mon.Not.R.Astr.Soc., 323, 453, \dodoi{10.1046/j.1365-8711.2001.04214.x}

\bibitem[{{Juri{\'c}} {et~al.}(2008){Juri{\'c}}, {Ivezi{\'c}}, {Brooks},
  {Lupton}, {Schlegel}, {Finkbeiner}, {Padmanabhan}, {Bond}, {Sesar},
  {Rockosi}, {Knapp}, {Gunn}, {Sumi}, {Schneider}, {Barentine}, {Brewington},
  {Brinkmann}, {Fukugita}, {Harvanek}, {Kleinman}, {Krzesinski}, {Long},
  {Neilsen}, {Nitta}, {Snedden}, \& {York}}]{Juric_etal_2008}
{Juri{\'c}}, M., {Ivezi{\'c}}, {\v{Z}}., {Brooks}, A., {et~al.} 2008,
  Astrophys.J., 673, 864, \dodoi{10.1086/523619}

\bibitem[{{Kafle} {et~al.}(2014){Kafle}, {Sharma}, {Lewis}, \&
  {Bland-Hawthorn}}]{Kafle_etal_2014}
{Kafle}, P.~R., {Sharma}, S., {Lewis}, G.~F., \& {Bland-Hawthorn}, J. 2014,
  Astrophys.J., 794, 59, \dodoi{10.1088/0004-637X/794/1/59}

\bibitem[{{Kalberla} \& {Kerp}(2009)}]{Kalberla+Kerp_2009}
{Kalberla}, P. M.~W., \& {Kerp}, J. 2009, Ann.Rev.Astron.Astrophys., 47, 27,
  \dodoi{10.1146/annurev-astro-082708-101823}

\bibitem[{{Khoperskov} {et~al.}(2020){Khoperskov}, {Gerhard}, {Di Matteo},
  {Haywood}, {Katz}, {Khrapov}, {Khoperskov}, \&
  {Arnaboldi}}]{Khoperskov_etal_2021}
{Khoperskov}, S., {Gerhard}, O., {Di Matteo}, P., {et~al.} 2020,
  Astron.Astrophys., 634, L8, \dodoi{10.1051/0004-6361/201936645}

\bibitem[{{Lin} \& {Li}(2019)}]{Lin+Li_2019}
{Lin}, H.-N., \& {Li}, X. 2019, Mon.Not.R.Astr.Soc., 487, 5679,
  \dodoi{10.1093/mnras/stz1698}

\bibitem[{{L{\'o}pez-Corredoira} \&
  {Betancort-Rijo}(2021)}]{Lopez-Corredoira+Betancort-Rijo_2021}
{L{\'o}pez-Corredoira}, M., \& {Betancort-Rijo}, J.~E. 2021, Astrophys.J., 909,
  137, \dodoi{10.3847/1538-4357/abe381}

\bibitem[{{L{\'o}pez-Corredoira} {et~al.}(2020){L{\'o}pez-Corredoira},
  {Garz{\'o}n}, {Wang}, {Sylos Labini}, {Nagy}, {Chrob{\'a}kov{\'a}}, {Chang},
  \& {Villarroel}}]{Lopez-Corredoira_etal_2020}
{L{\'o}pez-Corredoira}, M., {Garz{\'o}n}, F., {Wang}, H.~F., {et~al.} 2020,
  Astron.Astrophys., 634, A66, \dodoi{10.1051/0004-6361/201936711}

\bibitem[{{L{\'o}pez-Corredoira} \& {Sylos
  Labini}(2019)}]{Lopez-Corredoira_Sylos-Labini_2019}
{L{\'o}pez-Corredoira}, M., \& {Sylos Labini}, F. 2019, Astron.Astrophys., 621,
  A48, \dodoi{10.1051/0004-6361/201833849}

\bibitem[{{Lucy}(1977)}]{Lucy_1977}
{Lucy}, L.~B. 1977, Astron.J., 82, 1013, \dodoi{10.1086/112164}

\bibitem[{{Macci{\`o}} {et~al.}(2008){Macci{\`o}}, {Dutton}, \& {van den
  Bosch}}]{Maccio_etal_2008}
{Macci{\`o}}, A.~V., {Dutton}, A.~A., \& {van den Bosch}, F.~C. 2008,
  Mon.Not.R.Astron.Soc., 391, 1940, \dodoi{10.1111/j.1365-2966.2008.14029.x}

\bibitem[{{McGaugh}(2016)}]{McGaugh_2016}
{McGaugh}, S.~S. 2016, Astrophys.J., 816, 42,
  \dodoi{10.3847/0004-637X/816/1/42}

\bibitem[{{McGaugh}(2019)}]{McGaugh_2019}
---. 2019, Astrophys.J., 885, 87, \dodoi{10.3847/1538-4357/ab479b}

\bibitem[{{McGaugh} {et~al.}(2016){McGaugh}, {Lelli}, \&
  {Schombert}}]{McGaugh_etal2016}
{McGaugh}, S.~S., {Lelli}, F., \& {Schombert}, J.~M. 2016, Physical Review
  Letters, 117, 201101, \dodoi{10.1103/PhysRevLett.117.201101}

\bibitem[{{McMillan}(2017)}]{McMillan_2017}
{McMillan}, P.~J. 2017, Mon.Not.R.Astr.Soc., 465, 76,
  \dodoi{10.1093/mnras/stw2759}

\bibitem[{{Milgrom}(1983)}]{Milgrom_1983}
{Milgrom}, M. 1983, Astrophys.J., 270, 365, \dodoi{10.1086/161130}

\bibitem[{{Miyamoto} \& {Nagai}(1975)}]{Miyamoto+Nagai_1975}
{Miyamoto}, M., \& {Nagai}, R. 1975, Pub.Astron.Soc.Jap., 27, 533

\bibitem[{{Monari} {et~al.}(2018){Monari}, {Famaey}, {Carrillo}, {Piffl},
  {Steinmetz}, {Wyse}, {Anders}, {Chiappini}, \&
  {Jan{\ss}en}}]{Monari_etal_2018}
{Monari}, G., {Famaey}, B., {Carrillo}, I., {et~al.} 2018, Astron.Astrophys.,
  616, L9, \dodoi{10.1051/0004-6361/201833748}

\bibitem[{{Mr{\'o}z} {et~al.}(2019){Mr{\'o}z}, {Udalski}, {Skowron}, {Skowron},
  {Soszy{\'n}ski}, {Pietrukowicz}, {Szyma{\'n}ski}, {Poleski}, {Koz{\l}owski},
  \& {Ulaczyk}}]{Mroz_etal_2019}
{Mr{\'o}z}, P., {Udalski}, A., {Skowron}, D.~M., {et~al.} 2019,
  Astrophys.J.Lett, 870, L10, \dodoi{10.3847/2041-8213/aaf73f}

\bibitem[{{Navarro} {et~al.}(1997){Navarro}, {Frenk}, \&
  {White}}]{Navarro_etal_1997}
{Navarro}, J.~F., {Frenk}, C.~S., \& {White}, S.~D.~M. 1997, Astrophys. J.,
  490, 493, \dodoi{10.1086/304888}

\bibitem[{{Palunas} \& {Williams}(2000)}]{Palunas+Williams_2000}
{Palunas}, P., \& {Williams}, T.~B. 2000, Astron.J., 120, 2884,
  \dodoi{10.1086/316878}

\bibitem[{{Pfenniger} {et~al.}(1994){Pfenniger}, {Combes}, \&
  {Martinet}}]{Pfenniger_etal_1994}
{Pfenniger}, D., {Combes}, F., \& {Martinet}, L. 1994, Astron.Astrophys., 285,
  79.
\newblock \doarXiv{astro-ph/9311043}

\bibitem[{{Plummer}(1911)}]{Plummer_1911}
{Plummer}, H.~C. 1911, Mon.Not.R.Astr.Soc., 71, 460,
  \dodoi{10.1093/mnras/71.5.460}

\bibitem[{{Pouliasis} {et~al.}(2017){Pouliasis}, {Di Matteo}, \&
  {Haywood}}]{Pouliasis_etal_2017}
{Pouliasis}, E., {Di Matteo}, P., \& {Haywood}, M. 2017, Astron.Astrophys.,
  598, A66, \dodoi{10.1051/0004-6361/201527346}

\bibitem[{{Revaz} {et~al.}(2009){Revaz}, {Pfenniger}, {Combes}, \&
  {Bournaud}}]{Revaz_etal_2009}
{Revaz}, Y., {Pfenniger}, D., {Combes}, F., \& {Bournaud}, F. 2009,
  Astron.Astrophys., 501, 171, \dodoi{10.1051/0004-6361/200809883}

\bibitem[{{Sancisi}(1999)}]{Sancisi_1999}
{Sancisi}, R. 1999, Astrophys.Sp.Sci., 269, 59, \dodoi{10.1023/A:1017007620742}

\bibitem[{{Sancisi}(2004)}]{Sancisi_2004}
{Sancisi}, R. 2004, in Dark Matter in Galaxies, ed. S.~{Ryder}, D.~{Pisano},
  M.~{Walker}, \& K.~{Freeman}, Vol. 220, 233.
\newblock \doarXiv{astro-ph/0311348}

\bibitem[{{Sanders}(2010)}]{Sanders_2010}
{Sanders}, R.~H. 2010, {The Dark Matter Problem: A Historical Perspective}

\bibitem[{{Scarpa}(2006)}]{scarpa06}
{Scarpa}, R. 2006, in American Institute of Physics Conference Series, Vol.
  822, First Crisis in Cosmology Conference, ed. E.~J. {Lerner} \& J.~B.
  {Almeida}, 253--265, \dodoi{10.1063/1.2189141}

\bibitem[{{Sellwood}(1999)}]{Sellwood_1999}
{Sellwood}, J.~A. 1999, in Astronomical Society of the Pacific Conference
  Series, Vol. 182, Galaxy Dynamics - A Rutgers Symposium, ed. D.~R. {Merritt},
  M.~{Valluri}, \& J.~A. {Sellwood}, 351.
\newblock \doarXiv{astro-ph/9903184}

\bibitem[{{Sellwood} \& {Carlberg}(1984)}]{Sellwood+Carlberg_1984}
{Sellwood}, J.~A., \& {Carlberg}, R.~G. 1984, Astrophys.J., 282, 61,
  \dodoi{10.1086/162176}

\bibitem[{{Sellwood} \& {Carlberg}(2014)}]{Sellwood+Carlberg_2014}
---. 2014, Astrophys.J., 785, 137, \dodoi{10.1088/0004-637X/785/2/137}

\bibitem[{{Sellwood} \& {Carlberg}(2019)}]{Sellwood+Carlberg_2019}
---. 2019, Mon.Not.R.Astron.Soc., 489, 116, \dodoi{10.1093/mnras/stz2132}

\bibitem[{{Sellwood} \& {McGaugh}(2005)}]{Sellwood+McGaugh_2005}
{Sellwood}, J.~A., \& {McGaugh}, S.~S. 2005, Astrophys.J., 634, 70,
  \dodoi{10.1086/491731}

\bibitem[{{Sofue}(2020)}]{Sofue_2020}
{Sofue}, Y. 2020, Galaxies, 8, 37, \dodoi{10.3390/galaxies8020037}

\bibitem[{{Swaters} {et~al.}(2012){Swaters}, {Sancisi}, {van der Hulst}, \&
  {van Albada}}]{Swaters_etal_2012}
{Swaters}, R.~A., {Sancisi}, R., {van der Hulst}, J.~M., \& {van Albada}, T.~S.
  2012, Mon.Not.R.Astr.Soc., 425, 2299,
  \dodoi{10.1111/j.1365-2966.2012.21599.x}

\bibitem[{{Sylos Labini} {et~al.}(2020){Sylos Labini}, {Pinto}, \&
  {Capuzzo-Dolcetta}}]{SylosLabini_etal_2020}
{Sylos Labini}, F., {Pinto}, L.~D., \& {Capuzzo-Dolcetta}, R. 2020, Phys.Rev.E,
  102, 042108, \dodoi{10.1103/PhysRevE.102.042108}

\bibitem[{{Wang} {et~al.}(2023){Wang}, {Chrob{\'a}kov{\'a}},
  {L{\'o}pez-Corredoira}, \& {Sylos Labini}}]{Wang_etal_2023}
{Wang}, H.-F., {Chrob{\'a}kov{\'a}}, {\v{Z}}., {L{\'o}pez-Corredoira}, M., \&
  {Sylos Labini}, F. 2023, Astrophys.J., 942, 12,
  \dodoi{10.3847/1538-4357/aca27c}

\bibitem[{{Watkins} {et~al.}(2019){Watkins}, {van der Marel}, {Sohn}, \&
  {Evans}}]{Watkins_etal_2019}
{Watkins}, L.~L., {van der Marel}, R.~P., {Sohn}, S.~T., \& {Evans}, N.~W.
  2019, Astrophys.J, 873, 118, \dodoi{10.3847/1538-4357/ab089f}

\end{thebibliography}

\end{document}